\documentclass[11pt,reqno]{amsproc}
\usepackage[colorinlistoftodos,shadow]{todonotes}
\usepackage{amsmath,amscd,amssymb}
\usepackage{cite}
\usepackage{color}
\usepackage{graphicx}
\usepackage{psfrag,epsfig}
\usepackage{multirow}
\usepackage{rotating}
\usepackage{fullpage}
\usepackage[letterpaper,margin=1.0in]{geometry}
\usepackage{subfigure}
\usepackage{enumerate}
\usepackage{comment}
\usepackage{lineno}
\usepackage{algorithmic}
\usepackage{algorithm}
\usepackage[small]{caption}
\usepackage[debug=false, colorlinks=true, pdfstartview=FitV, linkcolor=blue, citecolor=blue, urlcolor=blue]{hyperref}
\numberwithin{equation}{section}
\newcommand*\patchAmsMathEnvironmentForLineno[1]{%
  \expandafter\let\csname old#1\expandafter\endcsname\csname #1\endcsname
  \expandafter\let\csname oldend#1\expandafter\endcsname\csname end#1\endcsname
  \renewenvironment{#1}%
     {\linenomath\csname old#1\endcsname}%
     {\csname oldend#1\endcsname\endlinenomath}}%
\newcommand*\patchBothAmsMathEnvironmentsForLineno[1]{%
  \patchAmsMathEnvironmentForLineno{#1}%
  \patchAmsMathEnvironmentForLineno{#1*}}%
\AtBeginDocument{%
\patchBothAmsMathEnvironmentsForLineno{equation}%
\patchBothAmsMathEnvironmentsForLineno{align}%
\patchBothAmsMathEnvironmentsForLineno{flalign}%
\patchBothAmsMathEnvironmentsForLineno{alignat}%
\patchBothAmsMathEnvironmentsForLineno{gather}%
\patchBothAmsMathEnvironmentsForLineno{multline}%
}

\title[Estimating Failure in Brittle Materials using Graph Theory]{Estimating Failure in Brittle Materials using Graph Theory}
\author[M.~K.~Mudunuru et al.]{M.~K.~Mudunuru$^{1,*}$, N.~Panda$^{2}$, S.~Karra$^{1}$, G.~Srinivasan$^{3}$, 
V.~T.~Chau$^{1}$, E.~Rougier$^{1}$, A.~Hunter$^{3}$, and H.~S.~Viswanathan$^{1}$ \\
{\scriptsize $^{1}$Earth and Environmental Sciences Division, 
Los Alamos National Laboratory, Los Alamos, NM 87545.} \\
{\scriptsize $^{2}$Theoretical Division, Los Alamos National 
Laboratory, Los Alamos, NM 87545.} \\
{\scriptsize $^{3}$X-Computational Physics Division, Los Alamos National 
Laboratory, Los Alamos, NM 87545.} \\
}
\thanks{$^*$Corresponding author, \texttt{maruti@lanl.gov}}
\date{\today}
\begin{document}
\maketitle
%
%
%
%
\section*{ABSTRACT} 
In brittle fracture applications, failure paths, regions where the failure occurs and damage statistics, are some of the key quantities of interest (QoI). 
High-fidelity models for brittle failure that  accurately predict these QoI exist but are highly computationally intensive, making them infeasible to incorporate in upscaling and uncertainty quantification frameworks.
The goal of this paper is to provide a fast heuristic to reasonably estimate quantities such as failure path and damage in the process of brittle failure.
Towards this goal, we first present a method to predict failure paths under tensile loading conditions and low-strain rates. 
The method uses a $k$-nearest neighbors algorithm built on fracture process zone theory, and identifies the set of all possible pre-existing cracks that are likely to join early to form a large crack. 
The method then identifies zone of failure and failure paths using weighted graphs algorithms.
We compare these failure paths to those computed with a high-fidelity model called the Hybrid Optimization Software Simulation Suite (HOSS).
A probabilistic evolution model for average damage in a system is also developed that is trained using 150 HOSS simulations and tested on 40 simulations.
A non-parametric approach based on confidence intervals is used to determine the damage evolution over time along the dominant failure path.
For upscaling, damage is the key QoI needed as an input by the continuum models.
This needs to be informed accurately by the surrogate models for calculating effective modulii at continuum-scale.
We show that for the proposed average damage evolution model, the prediction accuracy on the test data is more than 90\%.
In terms of the computational time, the proposed models are $\approx \mathcal{O}(10^6)$ times faster compared to high-fidelity HOSS.
These aspects makes the proposed damage model attractive for upscaling damage from micro-scale models to continuum models.
The proposed method in this paper is limited to tensile loading conditions at low-strain rates.
This loading conditions correspond to a dominant fracture perpendicular to tensile direction.
The proposed method is not applicable for in-plane shear, out-of-plane shear, and higher strain rate loading conditions.
\newline
\newline
\textbf{Keywords:}~fracture, graph theory, failure paths, damage statistics, brittle failure, fracture process zone.
%

\section{INTRODUCTION}
\label{Sec:S1_DynGraph_Intro}
Brittle fracture is a complex phenomena determined by interaction among several microstructural features of the material under study.
These features include grain size, presence of pre-existing cracks or defects and/or pores, frictional characteristics, etc \cite{1998_Bazant_Planas,1981_Petersson_LIT}. 
When a brittle material containing pre-existing cracks is loaded, stresses concentrate around the crack tips \cite{2013_Brooks_MIT,2007_Vesely_etal_p111_p118}. 
When these cracks propagate they interact with other defects and cracks in their neighborhood.
This results in a complex state of stress near the crack tips.
Especially at lower strain rates, pre-existing cracks act as nucleation points for the generation of new cracks \cite{2012_Freiman_Mecholsky,2016_Lamon}. 
Nucleation, growth, and coalescence of pre-existing cracks results in the degradation of elastic properties in brittle materials, eventually causing catastrophic failure.
This is because the critical amount of damage before rapid and unassisted crack growth can be limited in brittle materials.
Moreover, damage accumulation also results in non-linear material behavior, a further challenge for the development of predictive models.
Due to the prevalence of brittle materials in several applications in geosciences (e.g., hydraulic fracturing, geothermal, etc) \cite{hyman2016understanding,mudunuru2017sequential,mudunuru2017regression,2016_Mudunuru_SGW}, infrastructure \cite{2012_Freiman_Mecholsky} and aerospace industry \cite{Noor_Zarchan}, fast and accurate models of brittle fracturing are needed.

Many constitutive models and numerical methods have been proposed in the literature that simulate processes such as crack initiation, crack propagation, and crack shielding in brittle materials. 
Foundational work by Griffith \cite{Griffith:1921} and Irwin \cite{Irwin:1957} explained how crack growth can be understood as a free energy balance between the surface energy generated by crack formation and the elastic energy released by the crack as it grows, and the crack tip stress intensity factors.
Shortly thereafter, Barenblatt \cite{Barenblatt:1962} and Dugdale \cite{Dugdale:1960} proposed the idea of a cohesive zone or fracture process zone (FPZ) preceding a crack tip.
These ideas have been extensively expanded on in the past decades.
The current computing power available has allowed for complex fracture models to be applied to a wide range of length scales \cite{deBorst:2002,Lisjak:2014}.
Of these, many continuum (or macro-scale) approaches have been developed to address crack evolution and the overall material response in large samples (cm-scale and larger) including finite element methods \cite{Zienkiewicz_etal_SeventhEdition_v1,2012_Mudunuru__Nakshatrala_IJNME_v89_p1144_p1170,2015_MunjizaBook} and boundary element methods \cite{portela1992the,georige1981two}. 
Continuum-based approaches \cite{xu2016material} assume that the computational domain can be treated as one continuous body.
However, the nature of brittle fracture, which often results from growth and coalescence of major flaws, can be extremely localized and therefore not amenable to standard homogenization approaches.
This often results in the emergence of a non-physical length scale within the numerical methods used to implement continuum models that can significantly impact the calculated material response \cite{deBorst:1993}.
Hence, these numerical methods cannot account for localized strains without further enhancement of the mathematical formulation \cite{deBorst:1993, Belytschko:2009}.  
Discrete element methods \cite{kruggel2007review,Lisjak:2014} are another class of modeling techniques available at the same length scale as continuum approaches.
With these approaches, material is modeled as a collection of discrete blocks or particles that can displace and rotate with respect to each other, and even completely separate or break apart.
Similarly, there are combined finite-discrete element methods (FDEM) that also have discrete material blocks, which can deform themselves due to further resolution of each block with finite elements \cite{1999_Munjiza_etal,2004_Munjiza,2011_MunjizaBook,2015_MunjizaBook,Lei2016,2018_Osthus}.
In the FDEM methods, the discontinuities of interest (e.g., micro-cracks) must be on similar scale to the computational domain.
Of all these approaches available in the fracture mechanics literature, we adopt the FDEM approach in this work.
FDEM merges many features of other methods listed above.
The high fidelity FDEM model used in this work that accounts for fracture propagation was developed in a multi-physics software tool called Hybrid Optimization Software Suite (HOSS) \cite{HOSS,HOSS1}.

Although, FDEM models have been shown to accurately model brittle fracture, the use of these models (e.g., HOSS) in realistic simulations is computationally intractable as they resolve all the individual cracks with highly resolved meshes and small time-steps at large-scales. 
Despite the increased realism in the physics and parallel implementation, we are still unable to capture the full range of spatial or temporal scales due to the large computational requirements of our applications, such as hydraulic fracturing \cite{hyman2016understanding} or underground nuclear explosion monitoring \cite{jordan2015radionuclide}. 
Coarsening of the domain and simplification of the physics \cite{graham2013upscaling,kachanov2013introduction} are commonly used workarounds, but these methods eliminate critical topological features (critical effects of crack to crack interactions) and as a result, predictions routinely fail to match observations \cite{cady2012characterization,escobedo2014effect}. 
One then has to resort to upscaling methods where the domain is split into several grid cells and the material properties in each grid cell are obtained from QoI such as failure paths and damage statistics, which are still computationally challenging.
For example, to obtain the results for each HOSS simulation presented in this work, 4 hours of computation time on 400 processors were necessary. 
In addition, the amount of data being generated is also quite significant.
That is, each simulation took up to 11.5 GB of disk space to store all the information needed to characterize different fracture propagation stages.
Even for a very coarse upscaled domain of say 1000 cells, this leads to 1.6M CPU-hours and 11.5 TB of disk space.
For this reason, we need an approach that can provide key quantities of interest (QoIs) in a matter of seconds and with reasonable accuracy. 
This is the aim of the present work. We develop surrogate models for estimating the damage statistics along failure paths to inform continuum models. These surrogate models are built using data from HOSS high-fidelity numerical simulations.
The proposed methods presented in this work are applicable only under low-strain rates and Mode I failure or tensile loading conditions.
In these conditions, a single dominant fracture perpendicular to tensile loading is observed.
Applications where Mode I failure is important are hydraulic fracturing, material fracture toughness assessment, and structural integrity. 
A limitation of the proposed methodology is that it is not applicable for other modes of fracture and high strain rates loading conditions but the overall approach to build the surrogate models can be used.

In our previous work \cite{moore2018predictive}, we have utilized machine learning approaches to efficiently emulate the high fidelity model.
The key QoIs Moore et. al. \cite{moore2018predictive} focused on were time to failure and fracture coalescence predictions.
The previous work did not estimate damage, which is a key input needed for upscaling micro-scale information to macro-scale models.
Numerical codes such as FLAG \cite{fung2013ejecta,tonks2010mesoscale,tonks2012comparison} need information on how effective moduli degrade over time to simulate brittle damage at continuum-scales.
Current upscaling approaches in literature use empirical/phenomenological models based on experimental data or homogenization or statistical averaging techniques to inform damage to continuum models \cite{kachanov2013introduction}.
However, detailed micro-scale information about the damage evolution of the system is lost during this homogenization process \cite{krajcinovic1987micromechanics,krajcinovic1995some}.
The current work takes into account the detailed information available in HOSS simulations to account for damage evolution, which was not considered in previous works.
The proposed probabilistic damage evolution model constructed from HOSS simulations includes interaction between multiple cracks, which makes it robust for upscaling micro-scale models.
Our approach utilizes graphs to represent the topology of the system to capture failure paths that is then used to develop a reduced order model for damage along the failure path. 
We propose to overcome the computational cost hurdle associated with running high-fidelity dynamic fracture models by representing a fracture network as a graph, with far fewer DOFs ($\approx\mathcal{O}(10^4)$ times less), whose nodes and edges contain critical information about the topology and geometry of the cracks obtained from our high-fidelity HOSS model. 
Our new graph-based approach can be used to incorporate information such as damage statistics describing the evolution of the crack network, including the effects of crack-to-crack interactions, into the continuum codes seamlessly.
The main challenge in our work is computing the failure path using a weighted graph to represent crack growth.
Typically, in real materials, lengths, locations, and orientations of pre-existing cracks can be random.
The structure of the crack network may impact where the dominant flaw in the material forms.
Hence, one needs to account for this uncertainty in crack topology when modeling the brittle crack growth.

The main contribution of this study is to develop algorithms to predict failure paths and compute damage along the failure paths under tensile loading using weighted graphs at low-strain rates.
An advantage of the proposed method is that it is $\approx \mathcal{O} \left(10^6 \right)$ times faster than HOSS high-fidelity simulations.
Due to the increase in computational savings with reasonable predictions (accuracy of our damage model on unseen data is greater than 90\%), our methodology is ideal for usage in comprehensive uncertainty quantification studies which require 1000s of forward model runs.
The paper is organized as follows:~Sec.~\ref{Sec:S2_HOSS} details the set up for the 190 high-fidelity FDEM fracture propagation simulations using the HOSS simulator. This data set is then used for validating our proposed methods.
Sec.~\ref{Sec:S3_DynGraph_FailPaths} describes the proposed algorithm to estimate failure paths based on initial crack configuration.
The algorithm is based on a combination of FPZ theory, $k$NN, and Dijkstra's algorithms for weighted shorted paths.
Sec.~\ref{Sec:S4_DynGraph_Damage} details a method to construct and estimate damage evolution along the failure path.
Sec.~\ref{Sec:S5_DynGraph_Results} discusses the results of the proposed methods.
Failure path predictions are compared against HOSS simulations.
Details of training and validation of a damage evolution model are also provided in this section.
Prediction accuracy of the failure path algorithm and the damage evolution model are discussed as well.
Finally, conclusions are drawn in Sec.~\ref{Sec:S6_DynGraph_Conclusions}.
%

\section{HOSS SIMULATIONS AND FDEM MODEL FOR DYNAMIC FRACTURE}
\label{Sec:S2_HOSS}
HOSS is a FDEM code that was specifically designed to simulate problems involving fracture and fragmentation processes in a diverse range of materials \cite{HOSS}. 
Within the FDEM framework, finite element method solutions are combined with the discrete element method to simulate dynamic fracture. 
In this setting, solid domains are discretized using finite elements in order to describe their deformation and stresses.
Finite rotations and finite displacements are assumed \textit{a priori}.
That is, we do not make prior assumptions of small deformations or small rotations in our FDEM numerical formulation. 
To resolve fracture and contact processes between different parts, we allow material interaction along the boundaries of finite elements.
The finite element deformation kinematics is handled via a multiplicative decomposition-based finite strain formulation \cite{2015_MunjizaBook}.
The fracture model used in this present work is the combined single and smeared crack model introduced by Munjiza \cite{1999_Munjiza_etal}.
The contact interaction is resolved using the triangle-versus-point algorithm, which is the 2D extension of the tetrahedron-versus-point algorithm \cite{2011_MunjizaBook}. 

\begin{figure}
  \centering
  \includegraphics[width = 0.5\textwidth]
    {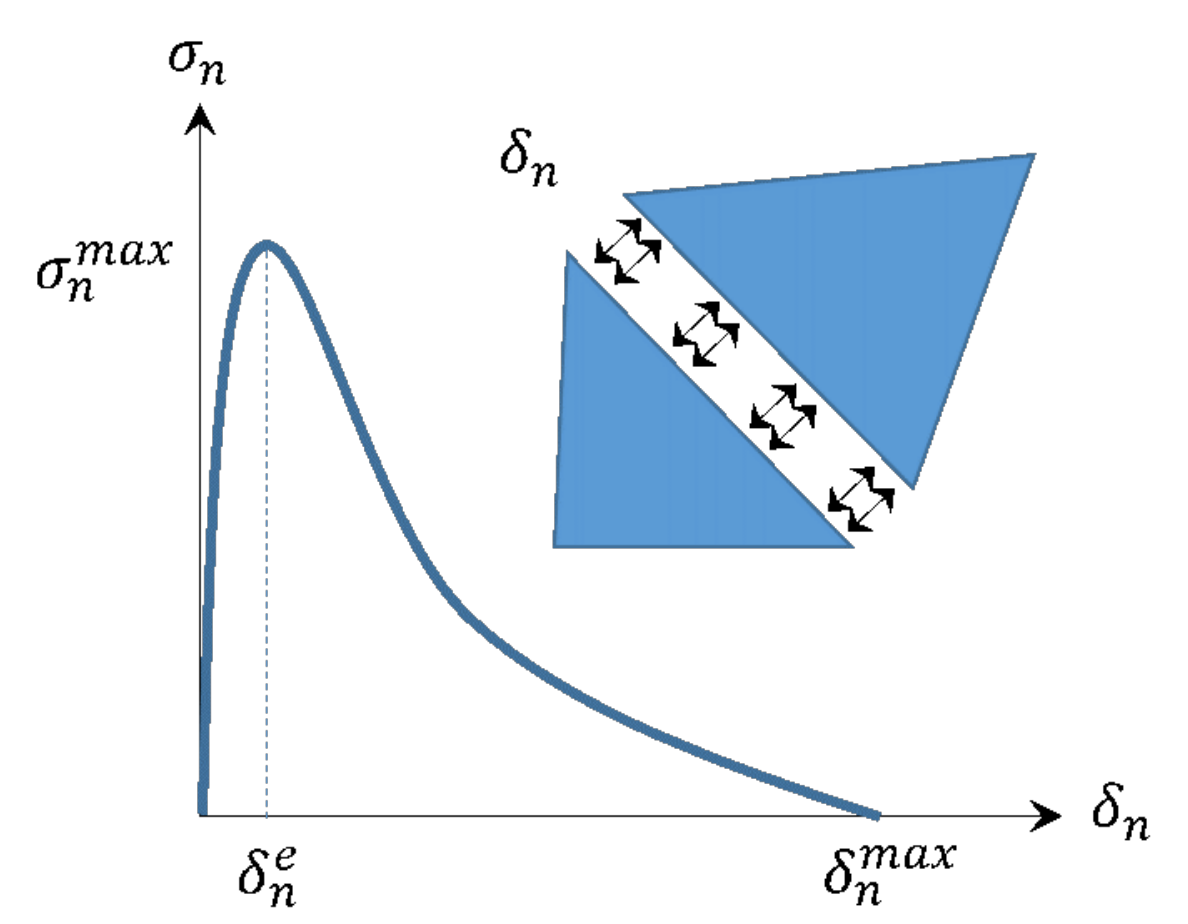}
  \caption{\textsf{\textbf{Fracture model:}}~A schematic representation of the fracture model in HOSS for Mode I loading.
  \label{Fig:FractureModel}}
\end{figure}

In the FDEM formulation, the interface between any two finite elements consists of non-linear springs that  model tensile and shear behavior. 
These springs can hold a maximum tensile stress equal to the tensile strength of the material $\sigma_n^{max}$ and a maximum shear stress that is based on a combination of the cohesion (Mohr-Coulomb model) and the frictional strength \cite{Labuz2012}.
Figure~\ref{Fig:FractureModel} presents a schematic representation of this curve. 
In the region $0 \leq \delta_n \leq \delta_n^e$ and $0 \leq \sigma \leq \sigma_n^{max}$, the springs undergo non-linear elastic behavior without any damage. 
Beyond the elastic limit, $\delta_n^e < \delta_n < \delta_n^{max}$, strain softening is assumed that mimics degradation in strength.
When $\delta_n > \delta_n^{max}$, these springs are broken and cannot bear any load.
The parameters in the strain softening are obtained from experiments on geomaterials \cite{2004_Munjiza}.

\begin{figure}
  \centering
  \includegraphics[width = 0.5\textwidth]
    {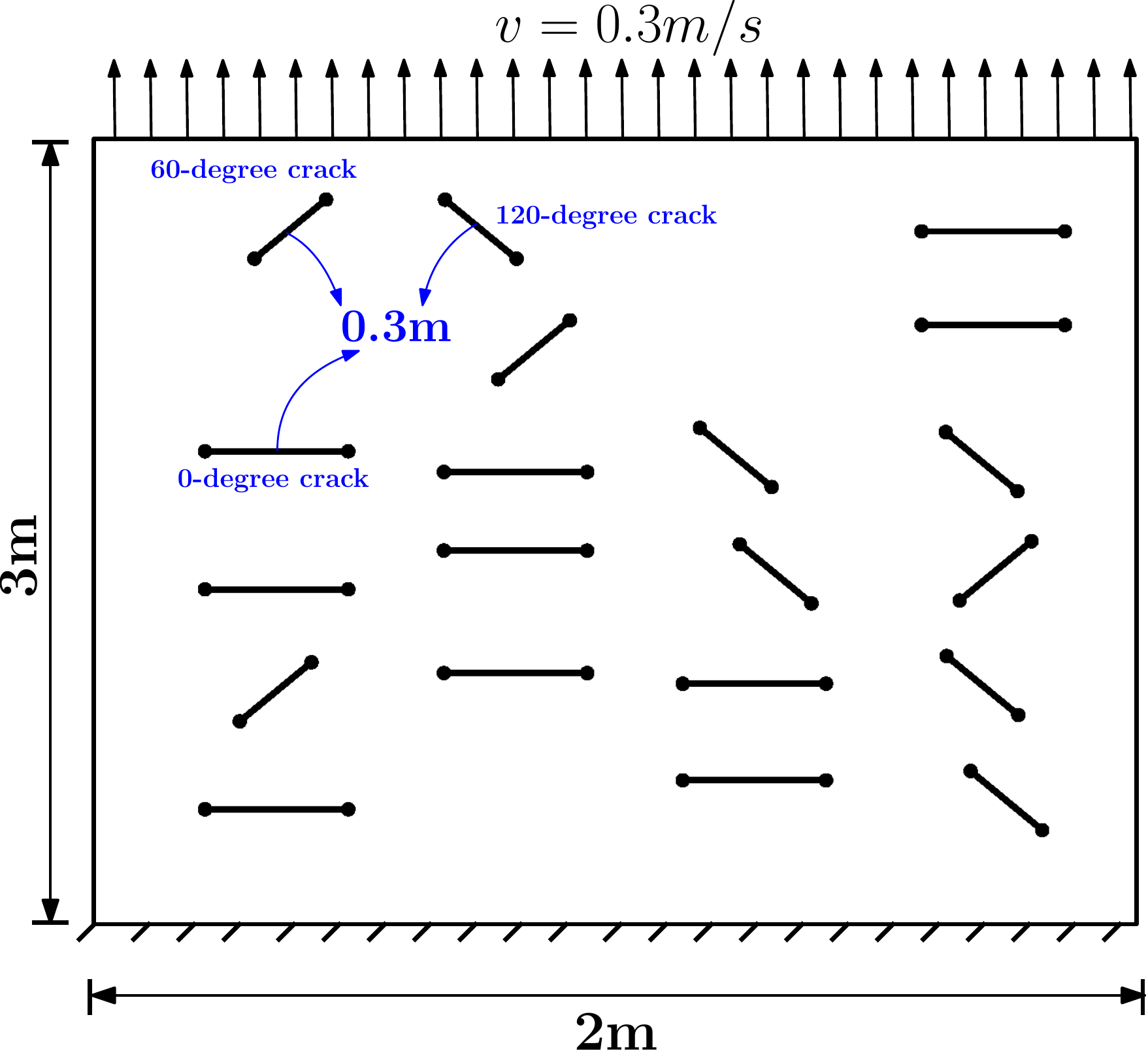}
  \caption{\textsf{\textbf{Initial boundary value problem:}}~A pictorial description of the HOSS simulation setup for a problem with 20 pre-existing cracks.
    190 high-fidelity HOSS simulations were performed. 
    For each simulation, the location and orientation of the initial cracks were randomly chosen.
    Orientations were chosen from 0-degree, 60-degree, or 120 degrees. 
    Loading conditions, material parameters, and domain dimensions are unchanged for all 190 HOSS simulations.
  \label{Fig:HOSS_20_Crack_Setup}}
\end{figure}

In this work, HOSS has been used to simulate brittle fracturing under uniaxial tensile loading conditions in concrete samples containing 20 randomly placed pre-existing cracks.
The sample size was set to 2 m wide and 3 m height.
A schematic of the numerical simulation setup is shown in Fig.~\ref{Fig:HOSS_20_Crack_Setup}. 
The bottom edge of the sample is fixed and the top edge of the sample moves upwards with a constant velocity of 0.3 m/s.
This boundary condition corresponds to a strain rate of 0.1 $\mathrm{s}^{-1}$. 
The lengths of pre-existing cracks was set to 0.3 m.
Their orientation was randomly selected between 0, 60, and 120 degrees with respect to horizontal.
Other parameters used include: sample density of 2500 kg/m$^3$, Young's modulus of 22.6 GPa, Shear modulus of 9.1 GPa, and a Poisson's ratio of 0.242.
The mesh used in this work contains constant strain triangles with an average size of 0.01 m.
This corresponds to a total of approximately 160,000 cells and 480,000 edges.
The tensile strength of the material was set to 8 MPa. 
Mohr-Coulomb fracture model is implemented at the interface of the triangular finite elements, which describes the strength of the material in shear.
The shear cohesion and the internal angle of friction were set at 24 MPa and 31 degrees respectively.
190 high-fidelity HOSS simulations were performed for different initial crack locations and orientations.
Loading conditions, material parameters, and domain dimensions are unchanged for all 190 HOSS simulations.
HOSS output was stored at every 2,000 time steps with each time-step being $10^{-8}$ seconds.
Each simulation took about 4 hours of computation time on 400 processors. 
At the point of failure, the material is unable to bear further load resulting in a catastrophic failure.
The sample is considered as failed when a set of fractures connect the two opposite boundaries of the domain.
In the next section, we present our method to estimate the failure path and its location.

\section{AN ALGORITHM TO ESTIMATE FAILURE PATHS}
\label{Sec:S3_DynGraph_FailPaths}
\begin{figure}
  \centering
  \includegraphics[width = 0.95\textwidth]
    {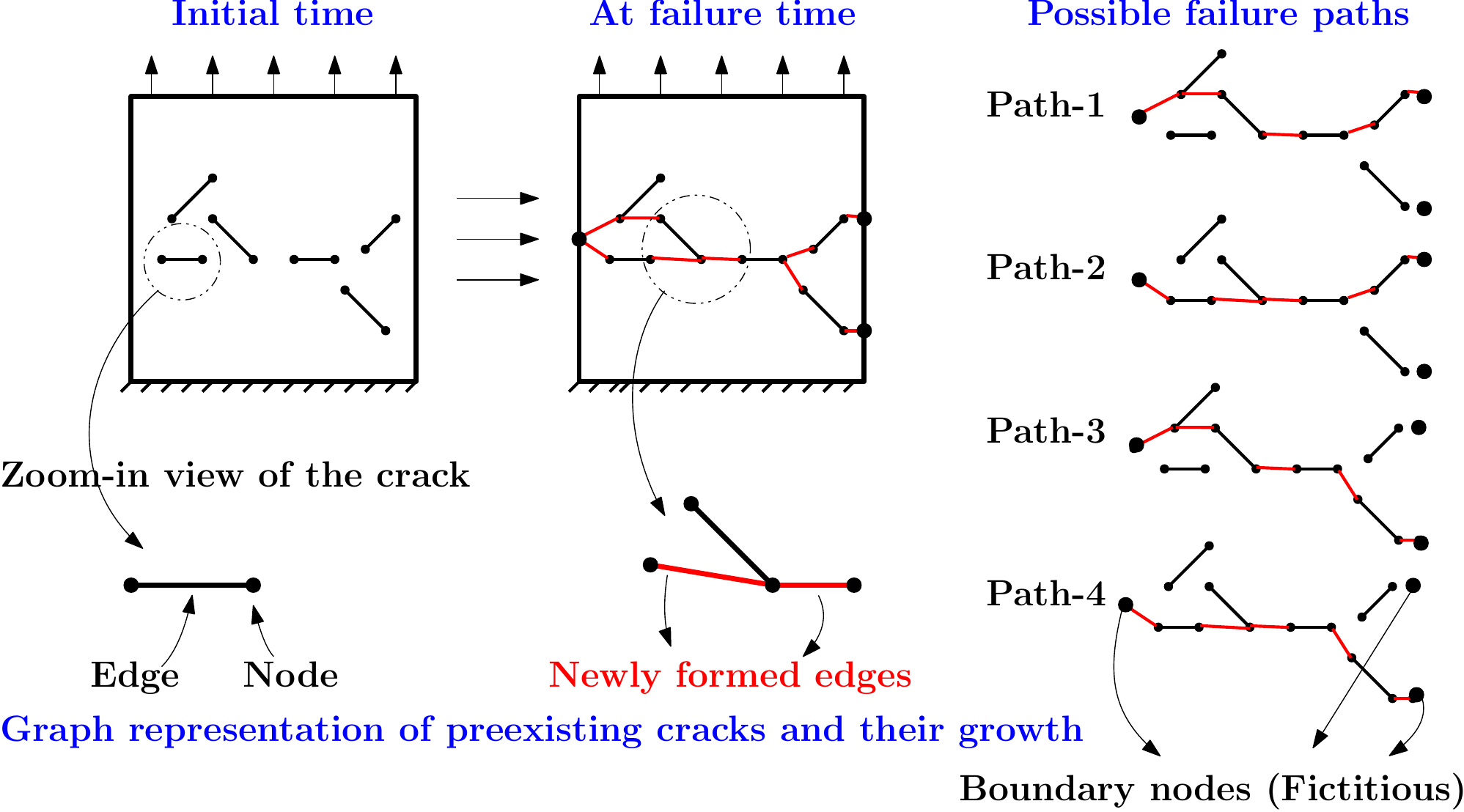}
  \caption{\textsf{\textbf{Failure paths graph representation:}}~A pictorial description of graph representation of fracture propagation.
  Crack tips are graph nodes and edges are cracks (which are assumed to be line segments connecting crack tips).
  \label{Fig:Graph_Rep_DynFrac}}
\end{figure}

In this section, we provide a detailed description of our proposed method to estimate failure paths.
The goal is to find the possible pathways to failure using weighted shortest path algorithms \cite[Chapter-3]{2013_Jungnickel} based on initial crack configuration.
Figure \ref{Fig:Graph_Rep_DynFrac} provides a graph based pictorial description of failure paths and Algorithm \ref{Algo:DFN_Length} summarizes the proposed method to estimate failure paths.
In our method, we assume graph nodes to be crack tips and graph edges are line segments connecting these tips.
At initial times, when the system is not loaded, our premise is that pre-existing cracks are not connected to each other.
In the graph theory representation, this implies that there is no connectivity between edges.
As the system is being loaded, pre-existing cracks grow and intersect with other cracks leading to the formation of new edges.
The schematic of all the possible paths connecting one side of the domain to another side at the time of failure for an initial pre-existing crack configuration is shown in Fig.~\ref{Fig:Graph_Rep_DynFrac}.
In fracture mechanics, there are three different modes of fracture:~Mode I, Mode II, and Mode III \cite{1990_Freund,1995_Unger,1998_Bazant_Planas}.
Mode I corresponds to tensile loading and is the most common load type encountered in fracture toughness testing of brittle materials \cite{2016_Lamon,1998_Krajcinovic_Vujosevic_IJSS_v35_p2487_p2503,2012_Freiman_Mecholsky}.
Mode II is in-plane shear and Mode III is out-of-plane shear or tearing mode.
The proposed method focuses on Mode I failure and while it is possible to extend the algorithm to consider other failure modes, it is outside the scope of this study.

Failure of brittle and quasi-brittle materials under Mode I typically starts with the development of a FPZ around the crack tip \cite{1981_Petersson_LIT,1992_Hu_Wittmann_v25_p319_p326,2013_Brooks_MIT}.
The FPZ is a region of high stress around the crack tip where damage accumulates as the crack propagates over time.
At a given instant of time, length of the FPZ is directly proportional to the length of the crack \cite{1995_Wang_etal_FFEMS}.
Very small micro-cracks are formed in the vicinity of the crack tip as stresses are the highest in this region.
As the crack advances over time, the micro-cracks within the FPZ coalesce to become a single entity.
This results in the formation of larger cracks.
Larger cracks are bound to have a stable crack growth compared to smaller cracks \cite{1990_Freund}.
Once these long cracks are formed, they grow rapidly with minimal additional loading.
Therefore, the possible failure paths correlates directly with FPZs, which is the basis for our proposed method.

The first step of the proposed method is to identify the cracks which are likely to coalesce.
In order to achieve this, we find pre-existing cracks that are orthogonal to the tensile loading.
These correspond to all $0$-degree angle cracks in the domain (as the orientation of these cracks is best suited for Mode I opening).
It is expected that they will have the fastest growth and we assume that the failure path will include one or more $0$-degree angle cracks. 
For each tip of these $0$-degree angle cracks, we calculate ten nearest crack tip neighbors using a $k$NN algorithm \cite{keller1985fuzzy} with $k=10$ and their respective Euclidean distances.
These nearest neighbor crack tips can belong to either $0$-degree, $60$-degree or a $120$-degree crack. 
Among these ten nearest crack tip neighbors, we identify the ones that could interact or coalesce with the horizontal cracks. 
Crack interaction or crack coalescence occurs if the FPZ of two neighboring cracks overlap. 
That is, if the Euclidean distance of a nearest neighbor is less than the length of FPZ, then we assume that the neighboring cracks are going to coalesce to form a larger crack.
In the graph based representation of failure, we form new edges by joining the crack tips that have overlapping FPZs.
The size of the FPZ, $\mathfrak{l}_{12}$, is given as \cite{1995_Wang_etal_FFEMS}:
\begin{align}
  \label{Eqn:Crack_Tip_Interaction}
  \mathfrak{l}_{12} \propto \left(\frac{\sigma}{\sigma_y} \right)^2 (l_1 + l_2)
\end{align}
where $\sigma$ and $\sigma_y$ are the applied and yield stresses of the material.
$l_1$ and $l_2$ are the length of the interacting cracks. 
Herein, we assume that the size of FPZ to be 75\% of $(l_1 + l_2)$. 

Once we have identified the potential coalescing cracks, we next identify the regions or zones of interest in which failure is likely to occur.
It should be noted that there may be one or more potential failure zones in a specimen.
However, in a realistic system only one of these failure paths will actually correspond to the sample's complete failure.
To identify this failure zone, we divide the entire domain into a set of non-overlapping rectangular zones.
Let the total number of non-overlapping rectangular zones be equal to \texttt{NumZones}.
The width of the rectangular zone is equal to the width of the domain and its height is equal to $\frac{H}{\texttt{NumZones}}$, where $H$ is the height of the specimen.
Then, we form a weighted undirected graph for the entire domain. 
This weighted graph contains graph nodes (which are crack tips), pre-existing edges, and newly formed edges.
Preexisting edges representing initial cracks are given small edge weights, equal to $10^{-4}$.
The rationale behind giving small edge weights to initial cracks is that, physically, there is a strong possibility for the failure path to traverse through these cracks.
The weights for newly formed edges are equal to their Euclidean distance.
Once this weighted graph is formed, we find all the connected components in each non-overlapping rectangular zone.
A connected component of a weighted undirected graph is a set of nodes such that each pair of nodes is connected by a path.
The Depth First Search algorithm available in \textsf{NetworkX} \cite{2008_NetworkX_Python} is used to identify connected components.
After identifying them, we search for the non-overlapping rectangular zone which has maximal connected component.
Here, the maximal connected component is defined as a connected component whose number of nodes is at least greater than or equal to the number of nodes in every other connected component in the graph.
The maximal connected component is identified in order of size, number of 0-degree cracks, largest length, and location with respect to loading side of the domain (see step-18 in Algorithm \ref{Algo:DFN_Length}).

In the final step of the proposed method, we look for weighted shortest paths connecting the sides of the domain which are parallel to the tensile loading direction. 
First, we introduce boundary nodes in the failure zone and create a weighted undirected graph within this zone.
New edges are constructed by connecting a given node to two of its nearest neighbors.
As mentioned previously, the weights for these newly formed edges are equal to the Euclidean distance and initial cracks are given relatively small edge weights.
Once a weighted graph is formed, we compute shortest paths from one boundary node to another boundary node.
Analysis is performed with and without the constraint that the path has to traverse through the identified maximal connected component.
Relaxing the above constraint gives more flexibility in identifying other possible paths (Sec.~\ref{Sec:S5_DynGraph_Results} discusses more on this aspect).
Dijkstra's algorithm \cite[Chapter-3]{2013_Jungnickel} is used to compute the weighted shortest paths.

\begin{algorithm} 
  \caption{{\small Failure paths prediction algorithm for many-crack geometries using weighted graphs}}
  \label{Algo:DFN_Length} 
  \begin{algorithmic}[1]
    \STATE INPUT:~Domain length; 
      Initial coordinates of the crack tips;
      \texttt{NumZones}:~Total number of regions the domain is divided into (in-order to estimate zone of failure);
      Length of FPZ.
    	\STATE Identify the 0-degree angle cracks that are perpendicular to tensile loading.
    \FOR{Each crack tip in the set of all 0-degree angle cracks:}
      \STATE Calculate ten nearest crack tip neighbors using $k$NN algorithm with $k=10$ and their respective Euclidean distances.
      \STATE Among these ten nearest crack tip neighbors, identify the ones that fall within the FPZ.
        These are determined as follows:
      \FOR{$i = 1, 2, \dots, 10$:}
        \IF{The Euclidean distance of an $i^{\mathrm{th}}$-nearest neighbor $\leq$ Length of FPZ:}
          \STATE Mark and store the $i^{\mathrm{th}}$-crack tip and the corresponding Euclidean distance.
        \ENDIF
      \ENDFOR
    \ENDFOR
    \STATE Form new edges by joining the crack tips that fall within the fracture process zone.
    \FOR{$i = 1, 2, \dots,$ \texttt{NumZones}:}
      \STATE Get and store the connected components in $i^{\mathrm{th}}$-zone.
      \STATE Get the total number of connected components, size of each connected component, and its length.
      \STATE Identify and store the 0-degree cracks present in these connected components.
    \ENDFOR
    \STATE Identify the failure zone. This is accomplished as follows:
      \begin{itemize}
        \item First, we get the largest/maximal connected component in each zone using the Depth First Search algorithm.
        \item Second, if two or more zones contain connected components that are of same size, then we choose a connected component which has maximum number of 0-degree cracks.
        \item Third, if the above set of connected components contain same number of 0-degree cracks, then we choose the one which has the largest length. 
        Length of a connected component is defined as the sum of edge weights.
        \item Finally, if they have the same length, then we choose the connected component which is closest to the loading side of the domain.
     \end{itemize}
    \STATE The goal is to detect failure paths along the length of sample, we introduce boundary nodes in the failure zone. 
      Their location is given as follows:
      \begin{itemize}
        \item If the index of the failure zone is `$i$', width of the domain is $L$, and $H$ is the height of the domain; then the boundary nodes are located at $(0,\frac{(i - 0.5) \times H}{\texttt{NumZones}})$ and $(L,\frac{(i - 0.5) \times H}{\texttt{NumZones}})$
      \end{itemize}
    \STATE Create a weighted graph within the failure zone. To achieve this we do the following:
      \begin{itemize}
        \item We search for two nearest neighbors for each crack tip. Then, we connect these two nearest neighbors with the corresponding crack tip to form new edges.
        \item The weights for these newly formed edges are equal to Euclidean distance. Initial cracks within the failure zone are given small edge weights = $10^{-4}$.
          This is because, physically, there is a strong possibility for the failure path to traverse through the pre-existing cracks.
        \item Once we construct the edges and their weights, we form a weighted graph.
      \end{itemize}
    \STATE Next, we compute shortest paths within this weighted graph using Dijkstra's method.
      The weighted shortest paths are calculated in two ways, both with and without the constraint that the path has to traverse through the identified connected component in the failure zone.
  \end{algorithmic}
\end{algorithm}
%

\section{AN ALGORITHM TO ESTIMATE DAMAGE}
\label{Sec:S4_DynGraph_Damage}
For upscaling to continuum-scale constitutive models, statistical information that describes damage accumulated over time is needed.
In this section, we provide an algorithm to estimate damage accumulation along failure paths from the HOSS simulation data.
As mentioned in Sec.~\ref{Sec:S2_HOSS}, within our FDEM framework, cracks are formed along the edges of the mesh elements.
For each pair of finite elements, we have a user specified number of non-linear springs.
For our simulations, we have assumed four normal and four shear springs whose behavior is shown in Fig.~\ref{Fig:FractureModel}.
As the crack grows, the interfacial springs are strained and irreversible damage is accrued as they enter a strain softening regime.
Damage accumulated per unit length between two finite elements is evaluated based on the strain of the springs that exceed $\delta_n^e$.
This value ranges from zero to one.
Zero damage value corresponds to undamaged springs ($\delta_n \leq \delta_n^e$) and damage value of one corresponds to completely broken springs $(\delta_n > \delta^{max}_n)$.

We use a non-parametric approach for determining the evolution of damage.
This is achieved by constructing a confidence interval over time. 
150 HOSS simulations are used for training and 40 simulations are used for testing the damage evolution model.
Let $t$ be the time index, our $95\%$ confidence interval for damage at $t$ is given by $\left(D^{(t)}_{L}, D^{(t)}_{U}\right)$.
$D^{(t)}_{L}$ and $D^{(t)}_{U}$ correspond to lower and upper estimate on damage at a given instance of time.
We use bootstrapping \cite{efron1994introduction,davison1997bootstrap} to estimate $D^{(t)}_{L}$, $D^{(t)}_{U}$, and mean damage $\overline{D^{(t)}_{M}}$ from our training data.
If a parametric distribution on the damage is desired, then one can prescribe a Gaussian distribution using the quantities, $\overline{D^{(t)}_{M}}$, $D^{(t)}_{L}$ and $D^{(t)}_{U}$ as follows, $D^{(t)}_{M} \sim \mathcal{N} \left(\overline{D^{(t)}_{M}}, \sigma^{(t)} \right)$.
Mean and variance of the distribution being $\overline{D^{(t)}_{M}}$ and $\sigma^{(t)}$, where $\sigma^{(t)} = \max\left(0.5* \left(\overline{D^{(t)}_{M}} - D^{(t)}_{L} \right), 0.5* \left(D^{(t)}_{U} - \overline{D^{(t)}_{M}} \right)\right)$.
Although we have not made it explicit, most damage occurs within our predicted path to failure. 
Of course other cracks will have some propagation, but in most cases failure is driven by a dominant fracture pathway.
For testing, to show that our confidence interval captures the damage from the test set, we calculate empirical coverage over time.
Empirical coverage represents the fraction of test cases that fall within our estimated confidence interval at anytime.
Mathematically, it is defined as, \[\frac{1}{N_T^{(t)}}\sum\limits_{j=1}^{N_T^{(t)}}\mathbb{I}_{D_j\in \left(D^{(t)}_{L}, D^{(t)}_{U}\right)},\]  where $N_T^{(t)}$ is the test set data at time $t$ and $D_j$ is one such data point of damage in our test set.
$\mathbb{I}_{D_j\in \left(D^{(t)}_{L}, D^{(t)}_{U}\right)}$ indicates if $D_j$ is in the confidence interval. 
The empirical coverage is one measure of how well our predicted confidence interval captures unseen data.
In the next section, we provide results and compare them with HOSS simulation data.
%

\section{RESULTS}
\label{Sec:S5_DynGraph_Results}
In this section, we present results of the proposed methods to estimate failure.
Following are the inputs given to Algorithm \ref{Algo:DFN_Length} to calculate failure paths: width of the domain is equal to 2 m, height of the domain is equal to 3 m,
\texttt{NumZones} = 3, initial crack tip coordinates, and length of FPZ = 0.45 m.
We use 190 simulations to test the proposed method to estimate failure path for the 20 crack problem.
In addition, we present a result for a configuration with 50 preexisting cracks (whose crack tip coordinates are chosen randomly) to demonstrate its predictive capability.

\subsection{Estimating failure paths}
\label{SubSec:S3_Path_to_Failure}
\begin{figure}
  \centering
  \subfigure[Cracks perpendicular to tensile loading (0-degree cracks)]
    {\includegraphics[width = 0.3\textwidth]
    {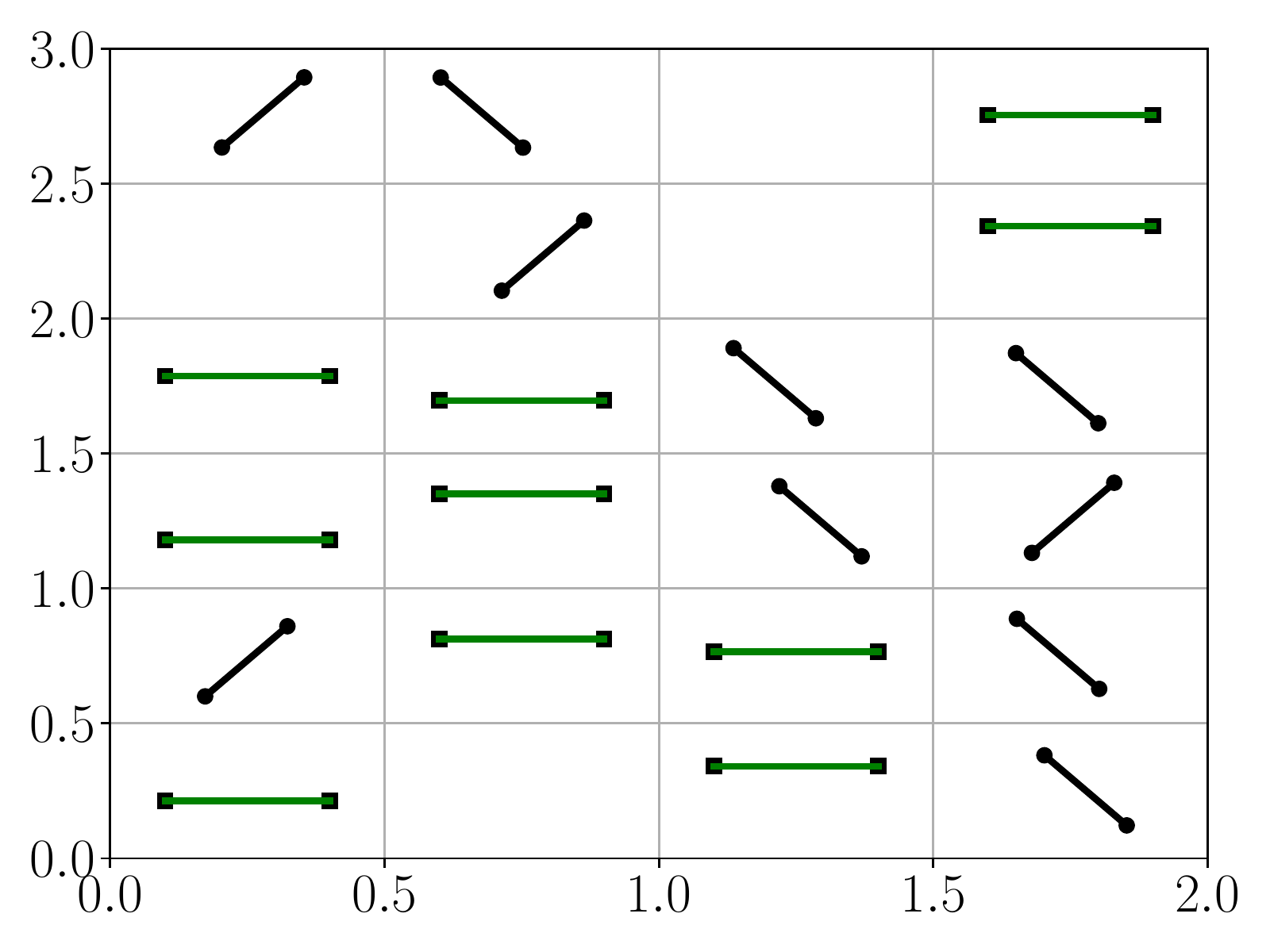}}
  \subfigure[Failure zone identification]
    {\includegraphics[width = 0.3\textwidth]
    {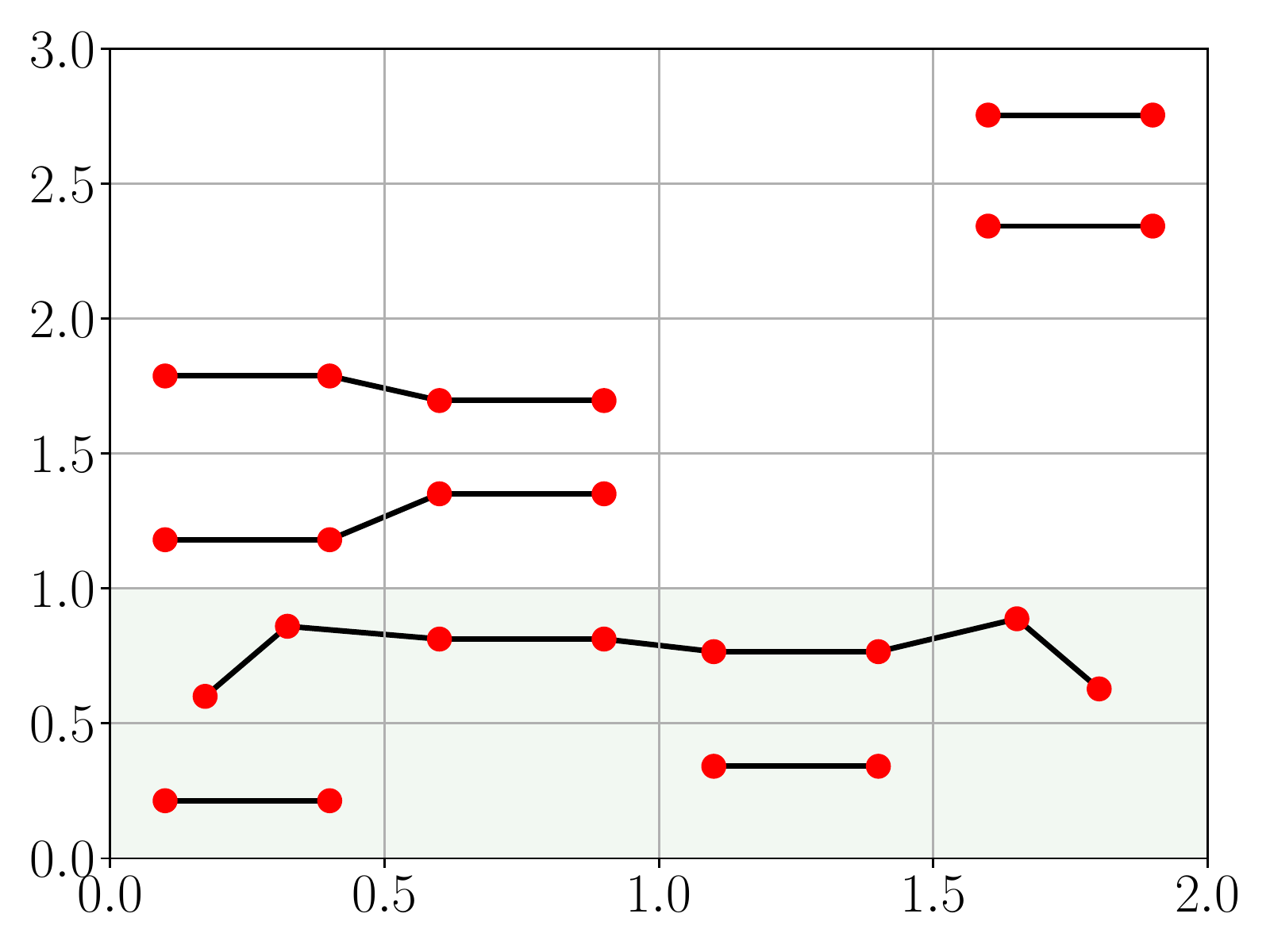}}
  \subfigure[Node-to-Node connectivity]
    {\includegraphics[width = 0.3\textwidth]
    {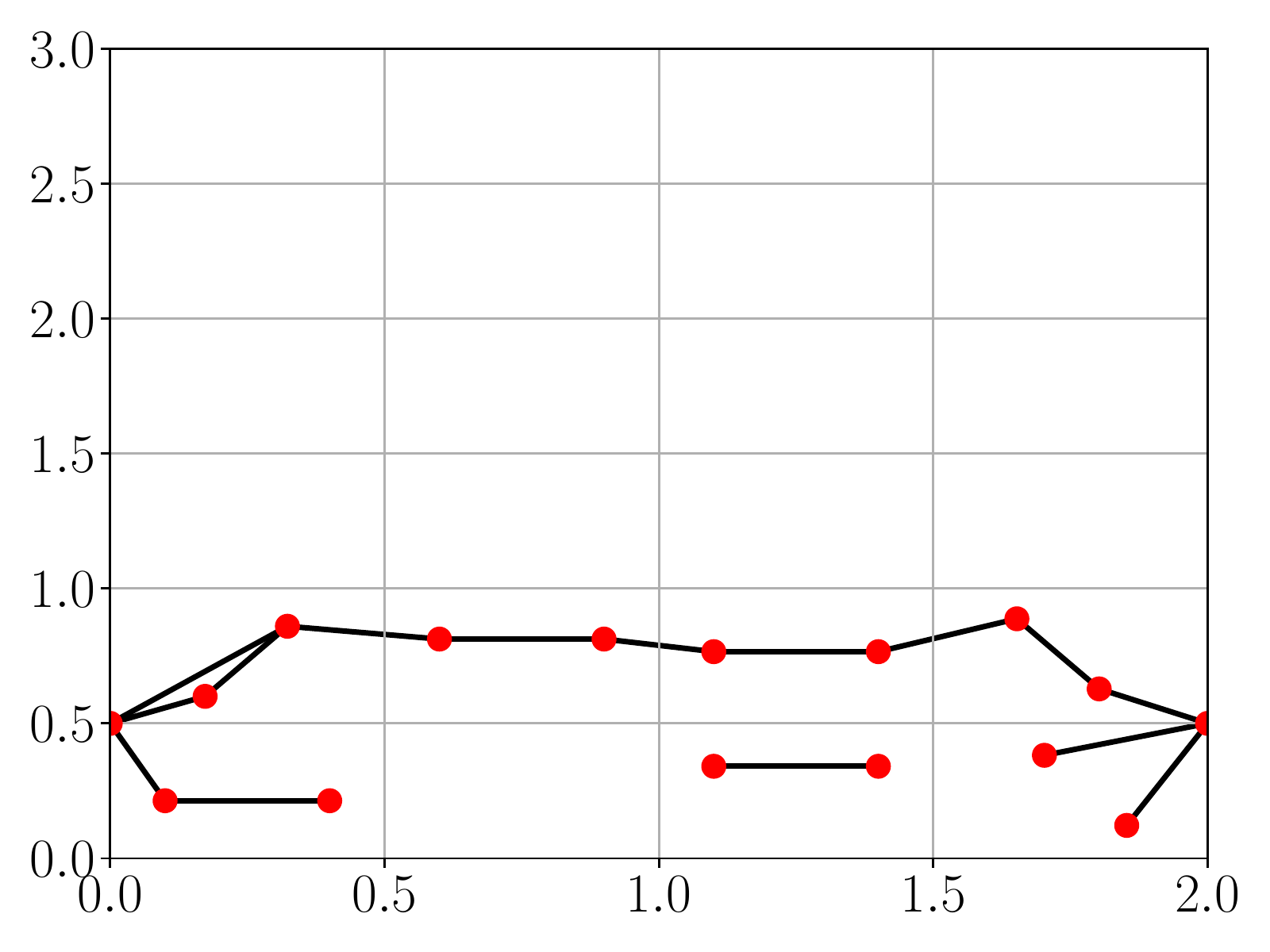}}
  \subfigure[Weighted graph based on 2-$NN$]
    {\includegraphics[width = 0.3\textwidth]
    {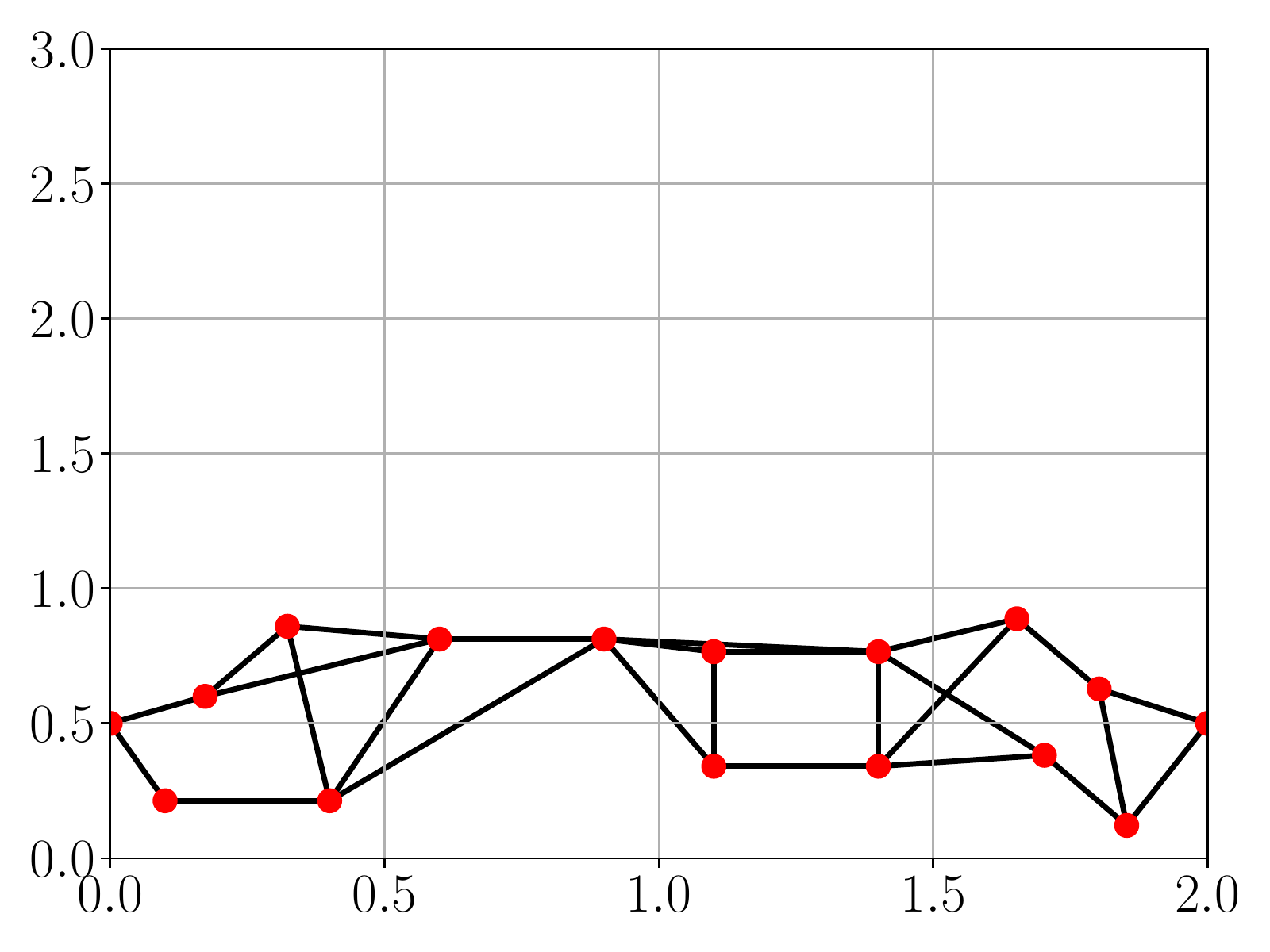}}
  \subfigure[Failure path-1 prediction]
    {\includegraphics[width = 0.3\textwidth]
    {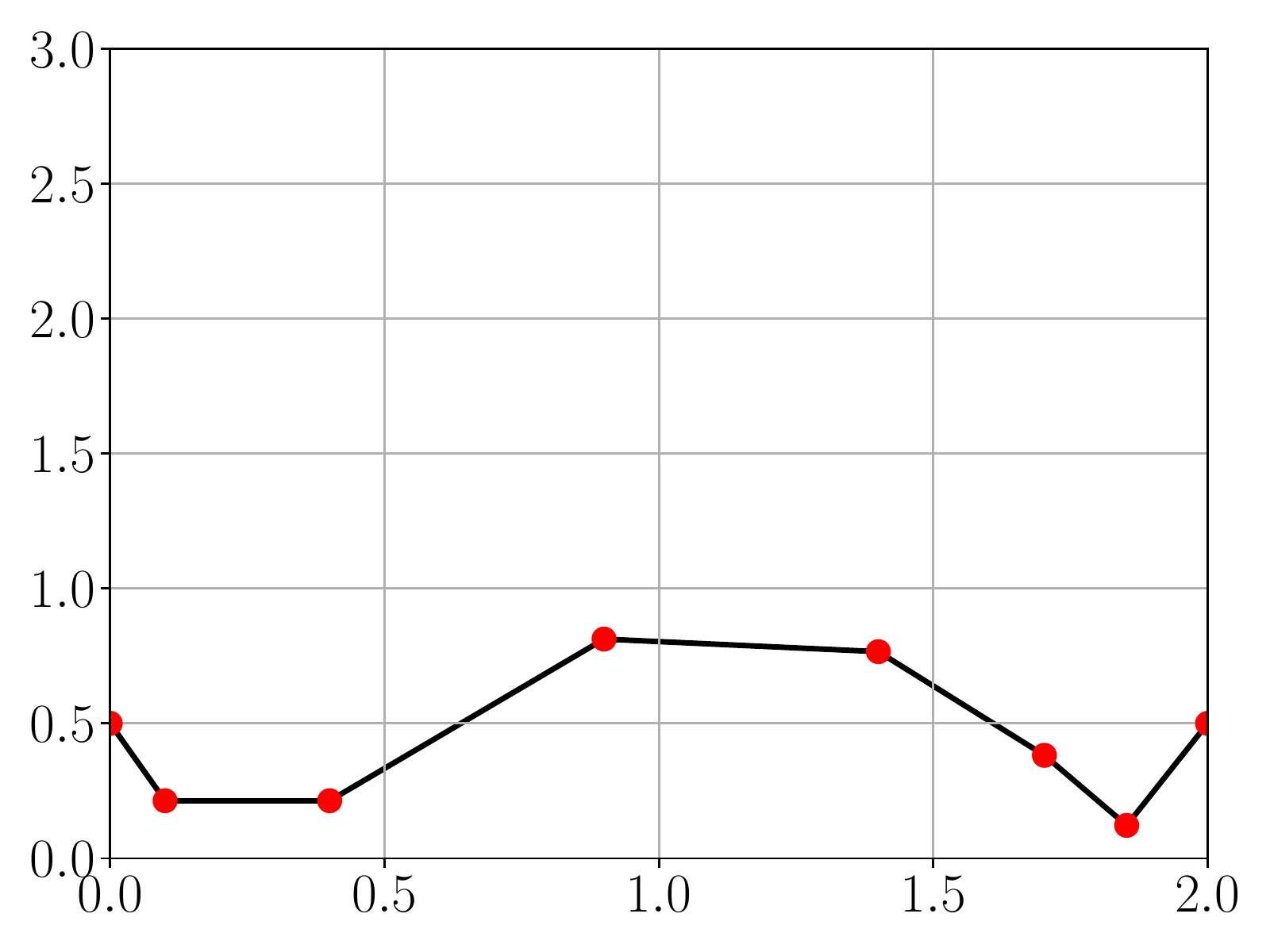}}
  \subfigure[Failure path-2 prediction]
    {\includegraphics[width = 0.3\textwidth]
    {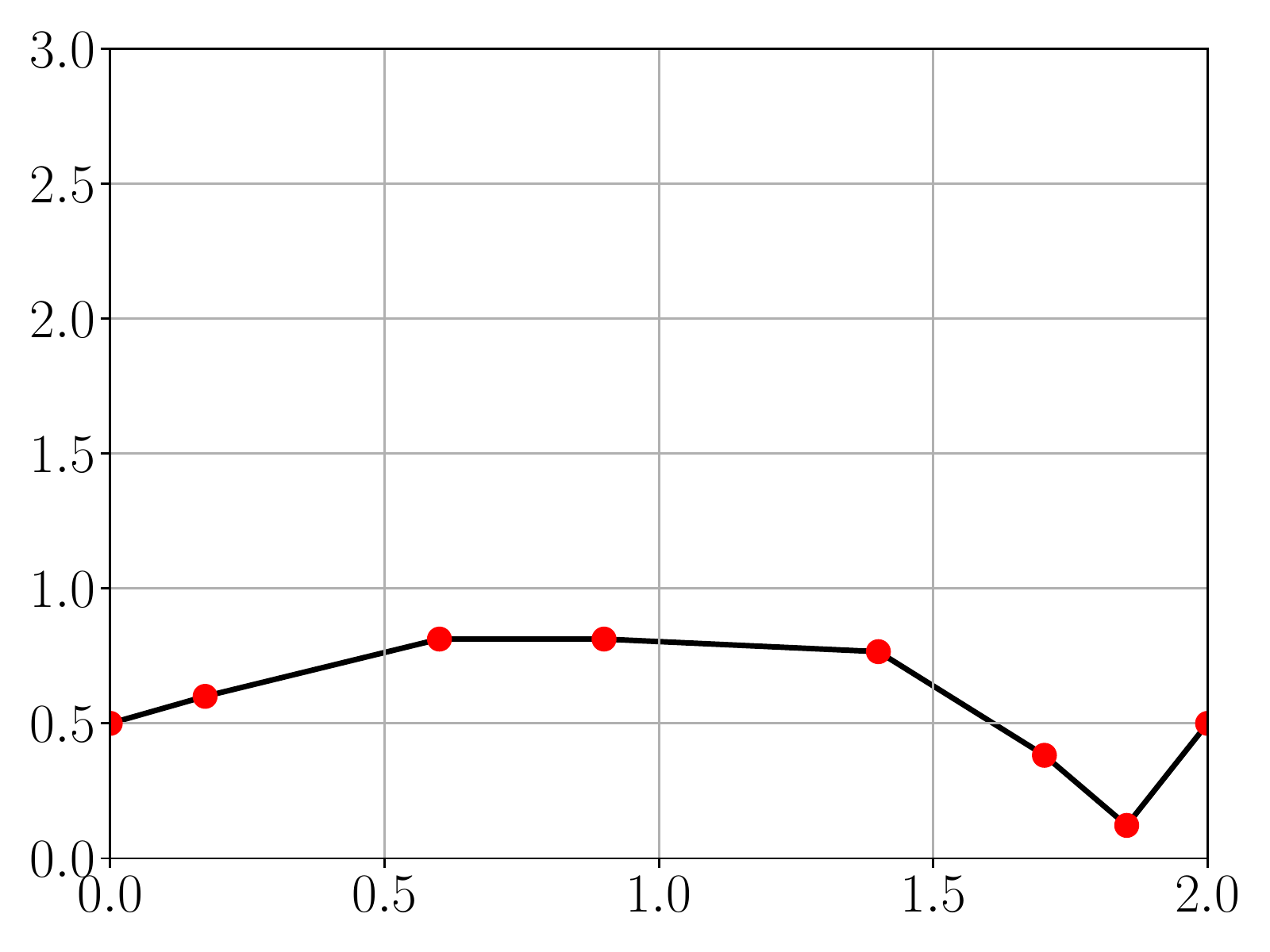}}
  \subfigure[Failure path-3 prediction]
    {\includegraphics[width = 0.3\textwidth]
    {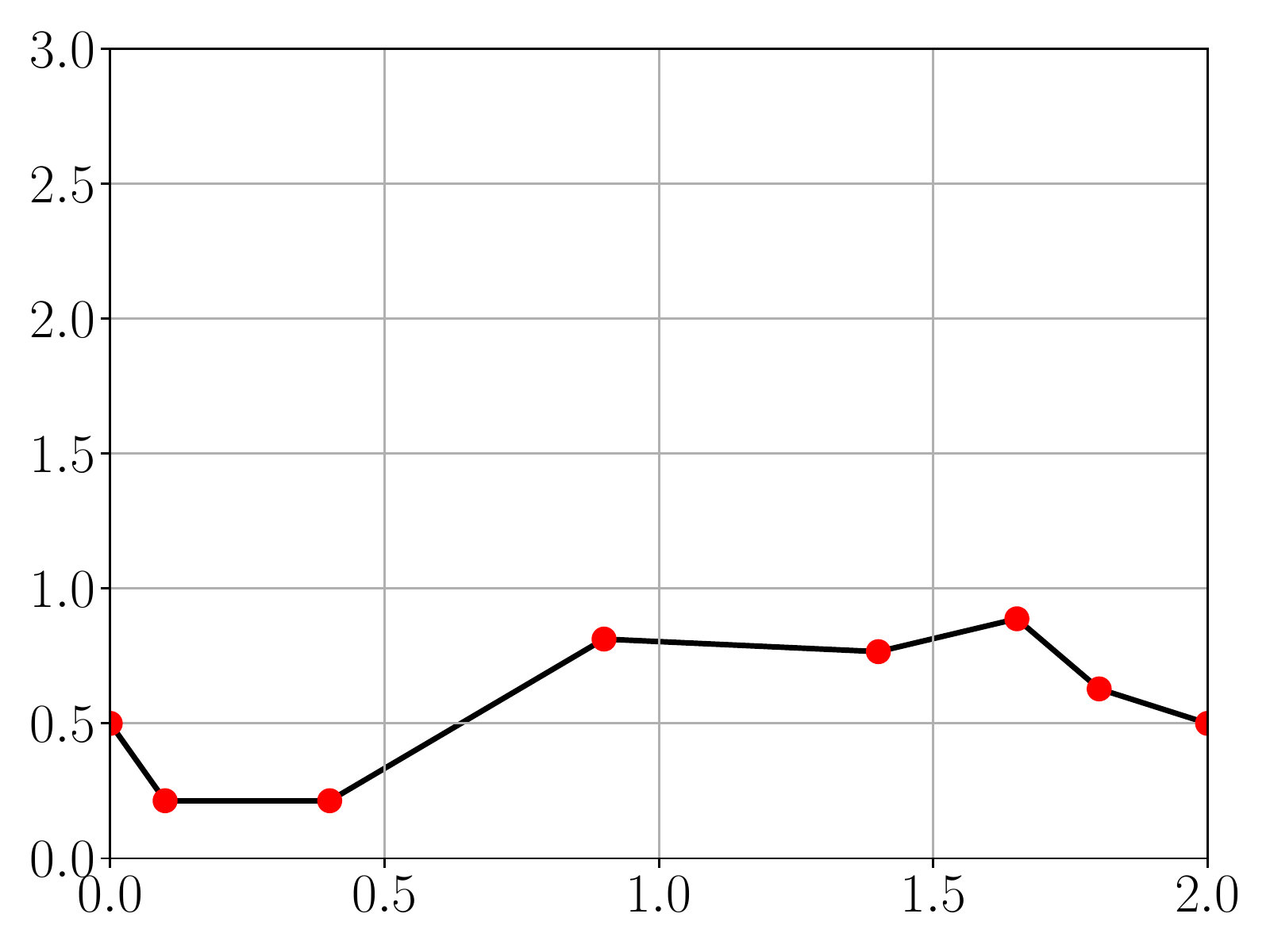}}
  \subfigure[Failure path-4 prediction]
    {\includegraphics[width = 0.3\textwidth]
    {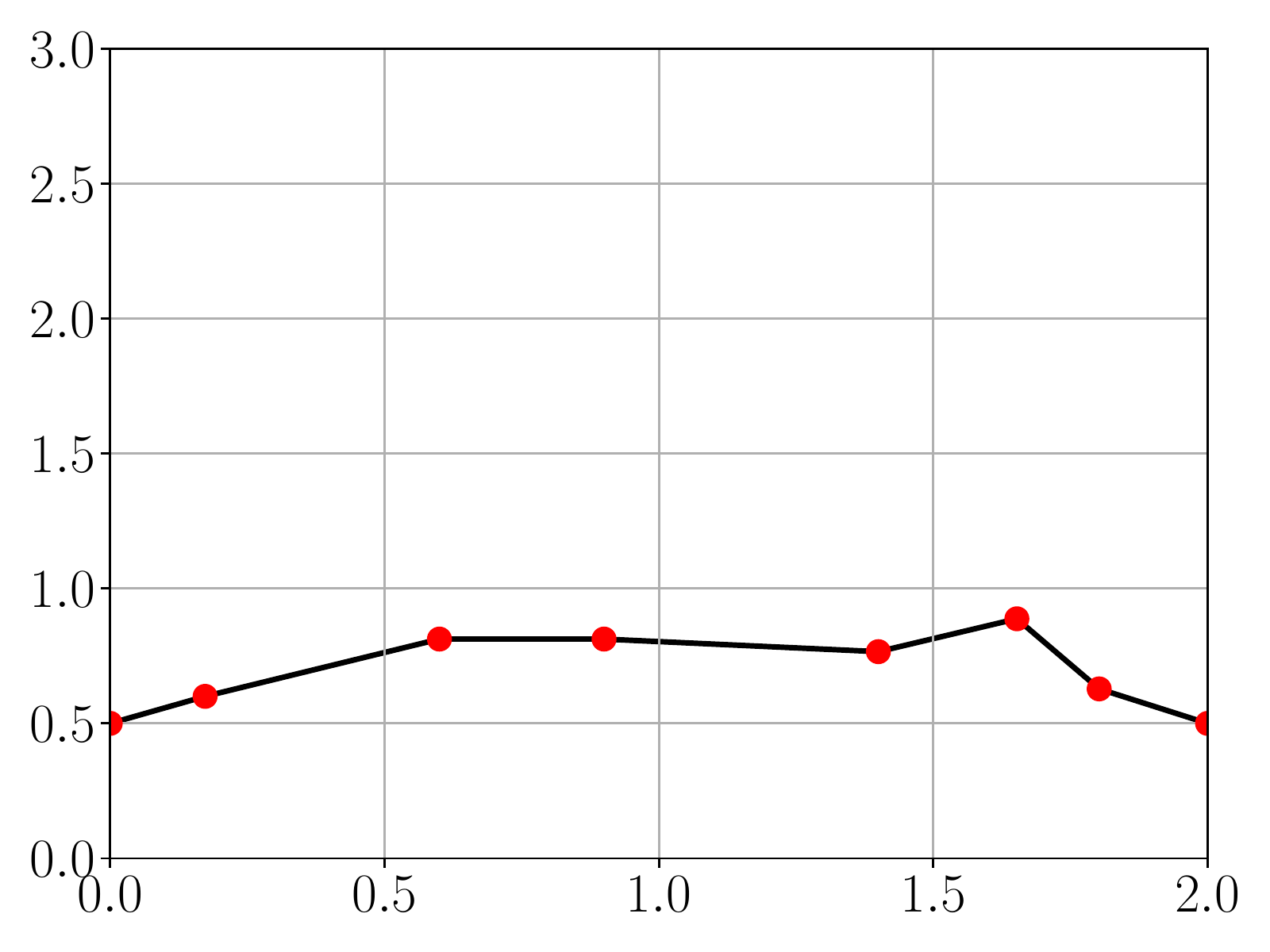}}
  \caption{\textsf{\textbf{Proposed method description:}}~The above figures show a step-by-step description of proposed method for estimating failure paths based on the method outline in Section \ref{Sec:S3_DynGraph_FailPaths}.
  \label{Fig:20_Cracks_Method_Descrip}}
\end{figure}

\begin{figure}
  \centering
  \includegraphics[width = 1.05\textwidth]
    {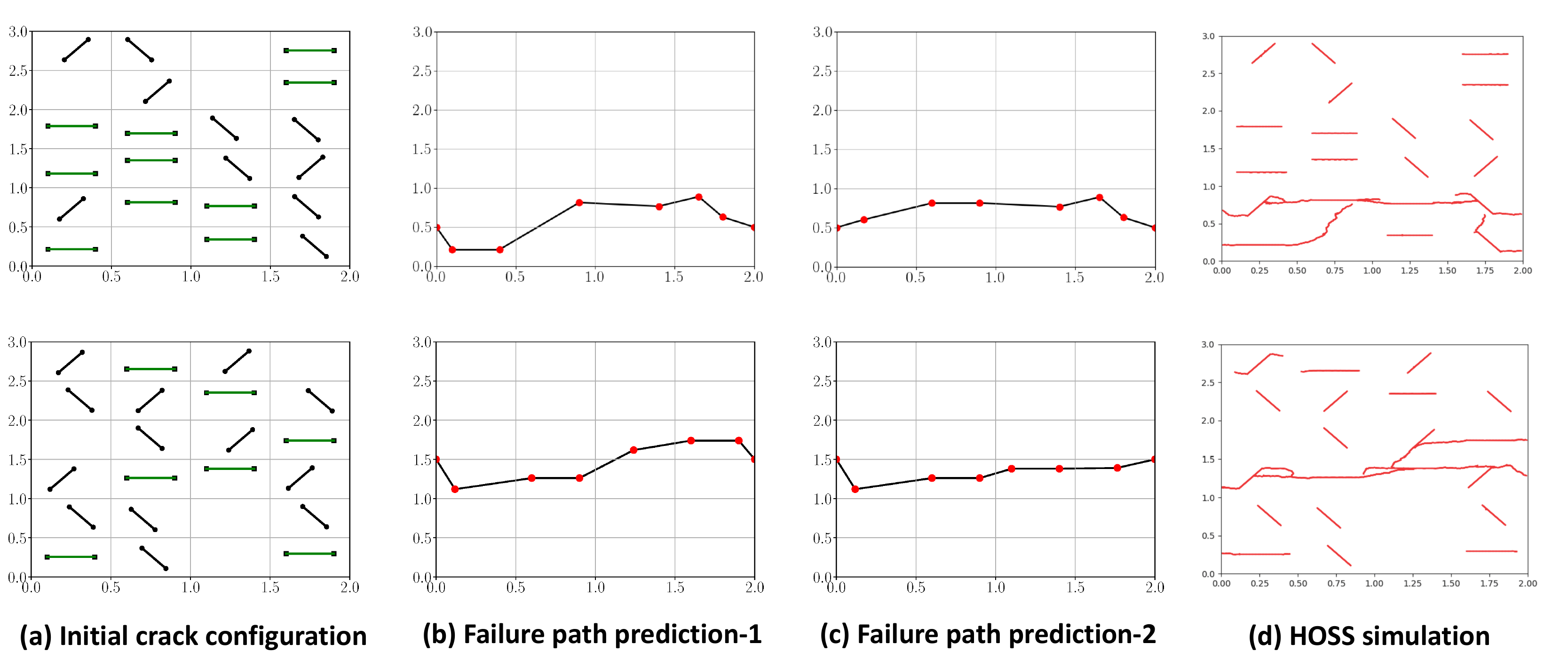}
  \caption{\textsf{\textbf{Failure path prediction (best scenarios, multiple predicted failure paths):}}~The above figures show the prediction of multiple failure paths using the proposed method.
  \label{Fig:NFPZ_20_Cracks_MultiplePaths}}
\end{figure}

Fig.~\ref{Fig:20_Cracks_Method_Descrip} provides a step-by-step description of the proposed method for estimating failure paths.
First, we identify cracks that are perpendicular to the tensile loading.
These are 0-degree cracks and are highlighted in green in Fig.~\ref{Fig:20_Cracks_Method_Descrip}(a).
Second, we identify zone of failure, which is shown in light blue in Fig.~\ref{Fig:20_Cracks_Method_Descrip}(b).
This corresponds to the region where failure occurs.
As described in Sec.~\ref{Sec:S3_DynGraph_FailPaths}, we identify larger cracks that are formed after connecting pre-existing cracks that fall within the FPZ to determine failure zone.
Third, we introduce boundary nodes in the failure zone (see Fig.~\ref{Fig:20_Cracks_Method_Descrip}(c)).
Fourth, we search for two nearest neighbors for each crack tip in the failure zone.
Then, we form new edges with edge weights being the Euclidean distance.
Fifth, we connect all the edges to form a weighted graph and look for failure paths connecting the boundary nodes (see Fig.~\ref{Fig:20_Cracks_Method_Descrip}(d)).
Finally, we compute all possible weighted shortest paths (see Fig.~\ref{Fig:20_Cracks_Method_Descrip}(e)-(h)).
The proposed method identified four possible shortest paths.
Among the four paths two paths are of exact match (see Fig.~\ref{Fig:NFPZ_20_Cracks_MultiplePaths}(d)) and other two are the next best match.
Failure paths are computed with and without enforcing the constraint that the shortest path has to traverse through the maximal connected component.
Herein, this constraint is enforced with some flexibility.
Meaning that, we find shortest paths that contain at least one node to be present in the maximal connected component.
This `soft' enforcement of the constraint is done with the intention to capture multiple failure paths (if they exist).

\begin{figure}
  \centering
  \includegraphics[width = 1.05\textwidth]
    {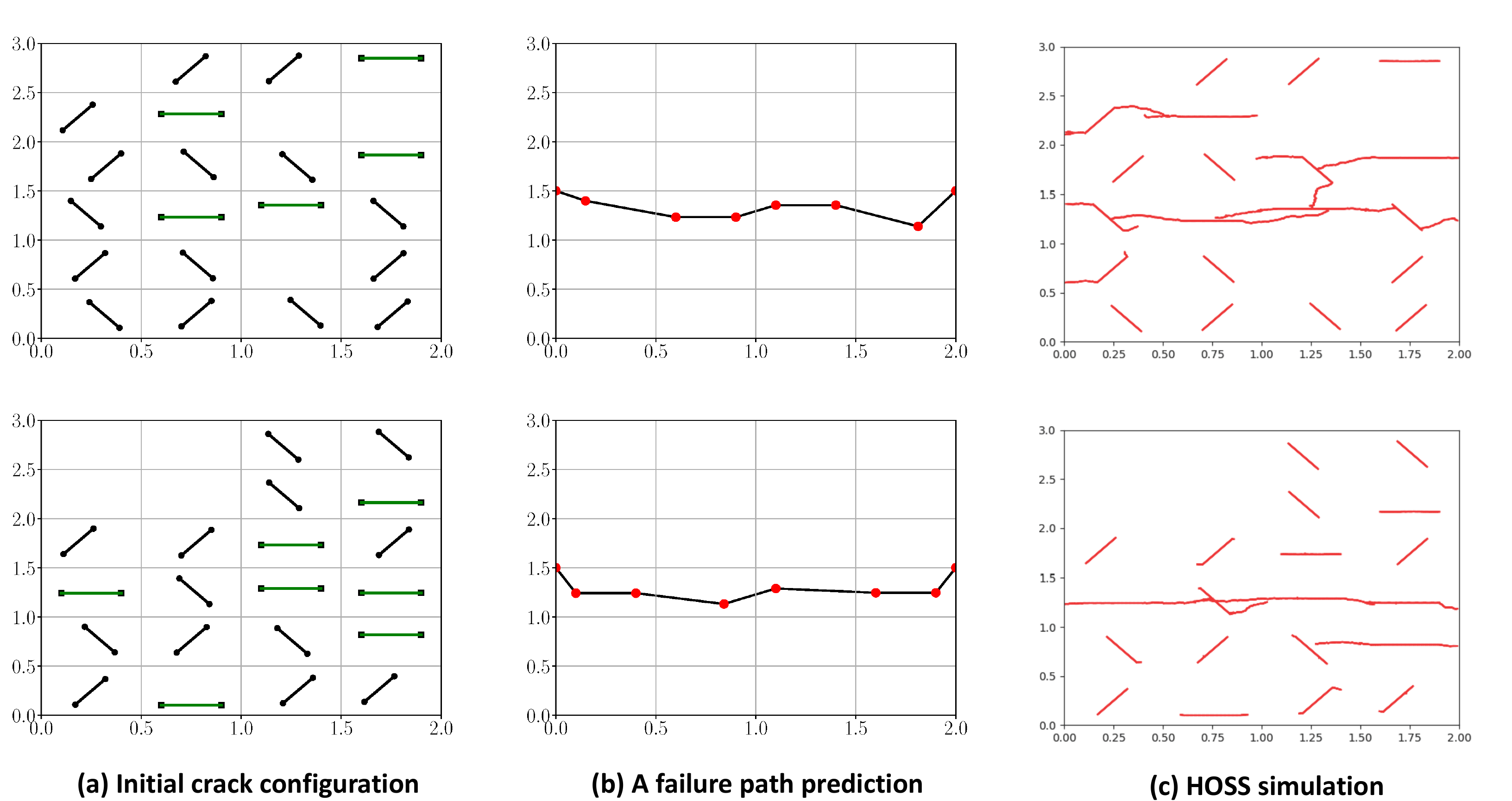}
  \caption{\textsf{\textbf{Failure path prediction (best scenarios, one failure path exact match):}}~The above set of figures show the prediction of failure paths with exact matched to the HOSS results.
  The left set of figures provide the initial crack configuration.
  The middle set of figures provide the prediction by the proposed method.
  The right set of figures show the HOSS results at failure.
  \label{Fig:NFPZ_20_Cracks_BestSinglePaths}}
\end{figure}

\begin{figure}
  \centering
  \includegraphics[width = 1.05\textwidth]
    {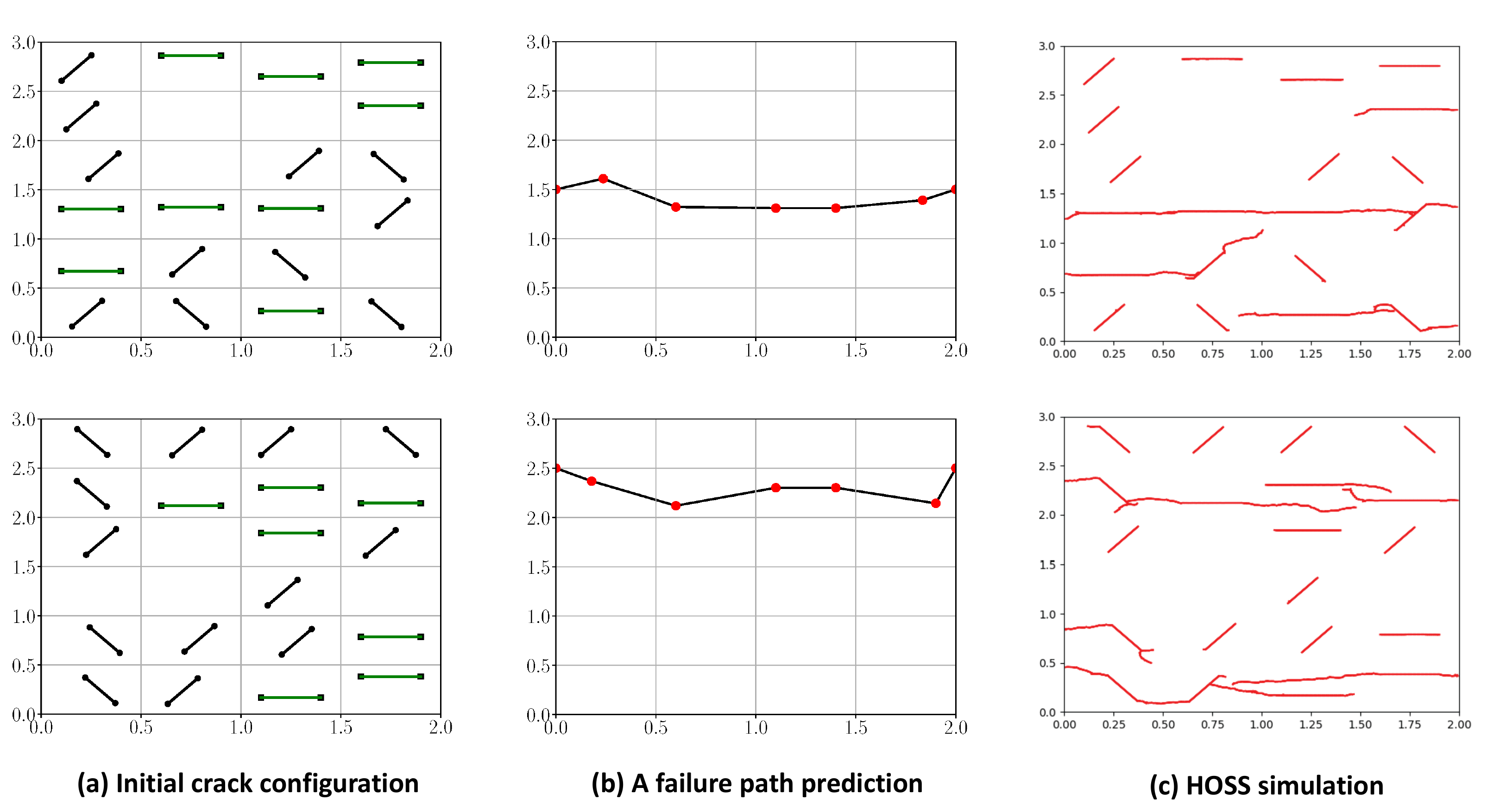}
  \caption{\textsf{\textbf{Failure path prediction (one failure path, greater than 50\% match):}}~The above set of figures show results where the initial cracks in the predicted failure path have a greater than 50\% match with those in the HOSS predicted failure path.
  \label{Fig:NFPZ_20_Cracks_NextBestPaths}}
\end{figure}

\begin{figure}
  \centering
  \includegraphics[width = 1.05\textwidth]
    {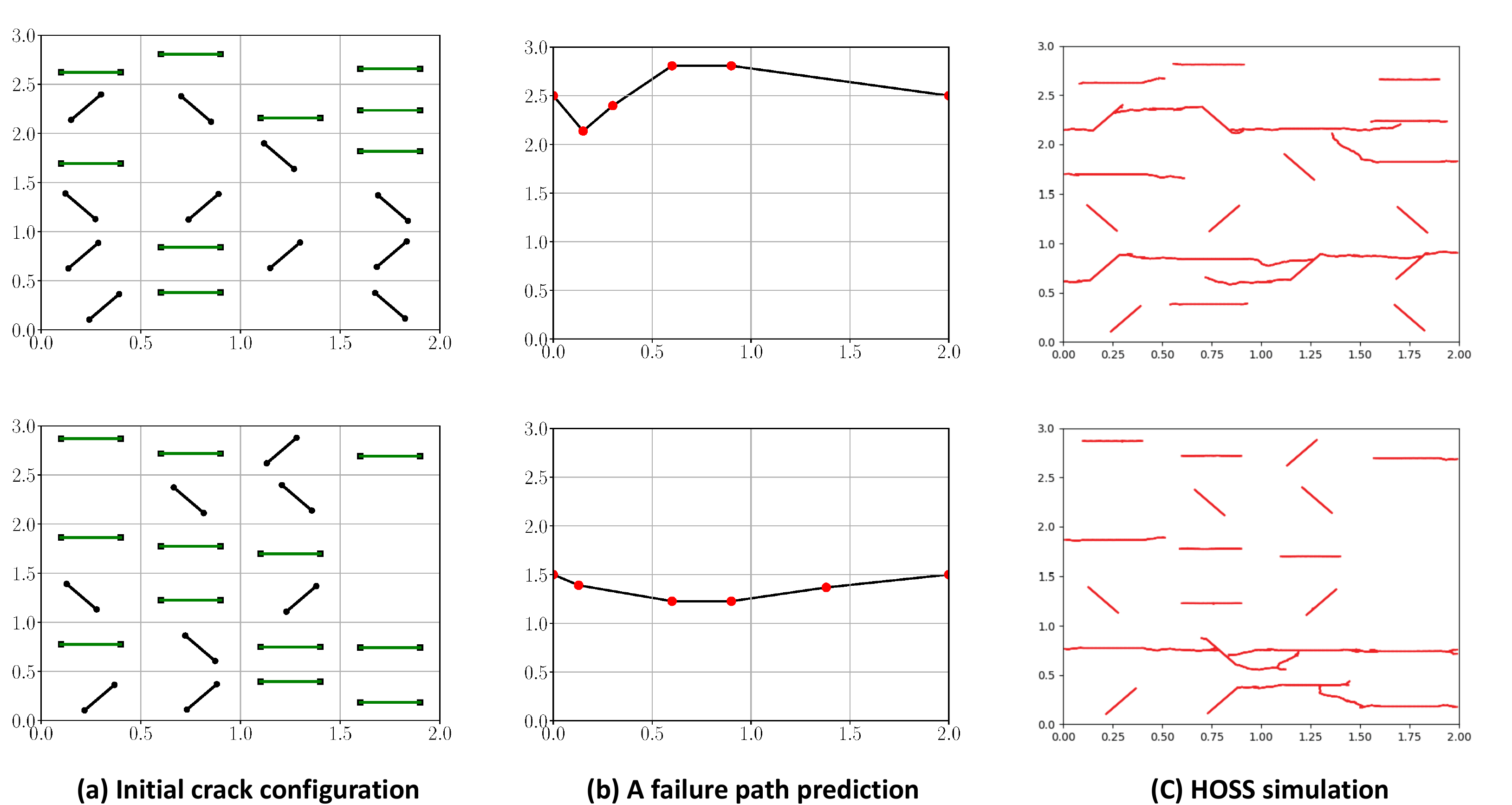}
  \caption{\textsf{\textbf{Failure path prediction (less than a 50\% match):}}~The above set of figures show the scenarios where the proposed method failed to predict the HOSS results.
  \label{Fig:NFPZ_20_Cracks_NonMatchPaths}}
\end{figure}

Figs.~\ref{Fig:NFPZ_20_Cracks_MultiplePaths}--\ref{Fig:NFPZ_20_Cracks_NonMatchPaths} show examples for failure path prediction and comparison with HOSS simulations.
When a possibility that multiple failure paths may exist, pinpointing an exact one is very hard.
Moreover, constructing a metric or criterion for an exact match can be challenging. 
Given an initial crack configuration, a plausible measure to assess the accuracy of the proposed method is as follows:~If the predicted failure path has atleast 50\% of the initial cracks or their crack tips to be within the HOSS failure path, then we say that the proposed method is a reasonable prediction of the simulated results.
Fig.~\ref{Fig:NFPZ_20_Cracks_MultiplePaths} presents example results in which the graph theory approach predicted two equally likely failure paths for one initial crack configuration.
Two different initial crack configurations are shown in Fig.~\ref{Fig:NFPZ_20_Cracks_MultiplePaths}(a).
The cracks perpendicular to tensile loading are highlighted in green.
Figs.~\ref{Fig:NFPZ_20_Cracks_MultiplePaths}(b) and (c) show the multiple failure paths predicted by the graphs for each case.
The HOSS predicted failure paths for both cases are provided in Fig.~\ref{Fig:NFPZ_20_Cracks_MultiplePaths}(d).
From these figures, it can be inferred that the proposed method is able to predict multiple failure paths for a given initial crack configuration. 

Fig.~\ref{Fig:NFPZ_20_Cracks_BestSinglePaths} shows example results from the graph theory approach for cases where there was only one predicted failure path with comparison to results from HOSS.
Fig.~\ref{Fig:NFPZ_20_Cracks_BestSinglePaths}(a) shows the initial crack configuration.
Fig.~\ref{Fig:NFPZ_20_Cracks_BestSinglePaths}(b) shows the failure path prediction using the proposed method.
Fig.~\ref{Fig:NFPZ_20_Cracks_BestSinglePaths}(c) shows the results obtained from the HOSS simulations.
It should be noted that it is not always possible to predict all paths of failure.
For example, the HOSS results shown in Fig.~\ref{Fig:NFPZ_20_Cracks_BestSinglePaths}(c) contain complex failure paths that could be considered two distinct pathways. 
However, our method provided only a single failure path.
There are 43  out of 190 simulations where the graph approach has predicted a single failure path and matched perfectly with atleast one failure path of HOSS simulation.
By perfect match, we mean that all of the initial cracks or crack tips in the graph predicted failure path are also in the HOSS predicted path.

Fig.~\ref{Fig:NFPZ_20_Cracks_NextBestPaths} shows example results from the graph theory approach for cases where there was only a partial match.
Quantitatively, by partial match we mean atleast 50\% of the initial cracks are in the failure paths in comparison to the HOSS simulations.
Predictions that have more than 50\% of the initial cracks in the failure paths and also have a zone of failure similar to that of HOSS are a total of 75 out of 190. 
Note that these 75 are separate from the previously mentioned 43 simulations. 
Fig.~\ref{Fig:NFPZ_20_Cracks_NonMatchPaths} shows example scenarios where the proposed method failed to match the HOSS results. 
Predictions that fall in this category are the remaining 72 out of 190.
Table \ref{Tab:NFPZ_Accuracy} summarizes the accuracy of the proposed method, which is the number of reasonably accurate predictions made by the Algorithm \ref{Algo:DFN_Length}.
Out of 190 simulation, the proposed method predicted the failure paths of 118 simulations with reasonable (50\% of the failure path) or accurate prediction (100\% of the failure path).
The prediction accuracy of failure zones by the proposed method is 143 simulations out of 190 simulations.
It should be noted that the time taken by the proposed method to predict a failure path is a couple of seconds, which is $\approx \mathcal{O}(10^6)$ times less than a single HOSS simulation.

\begin{figure}
  \centering
  \subfigure[Initial crack network]
    {\includegraphics[width = 0.375\textwidth]
    {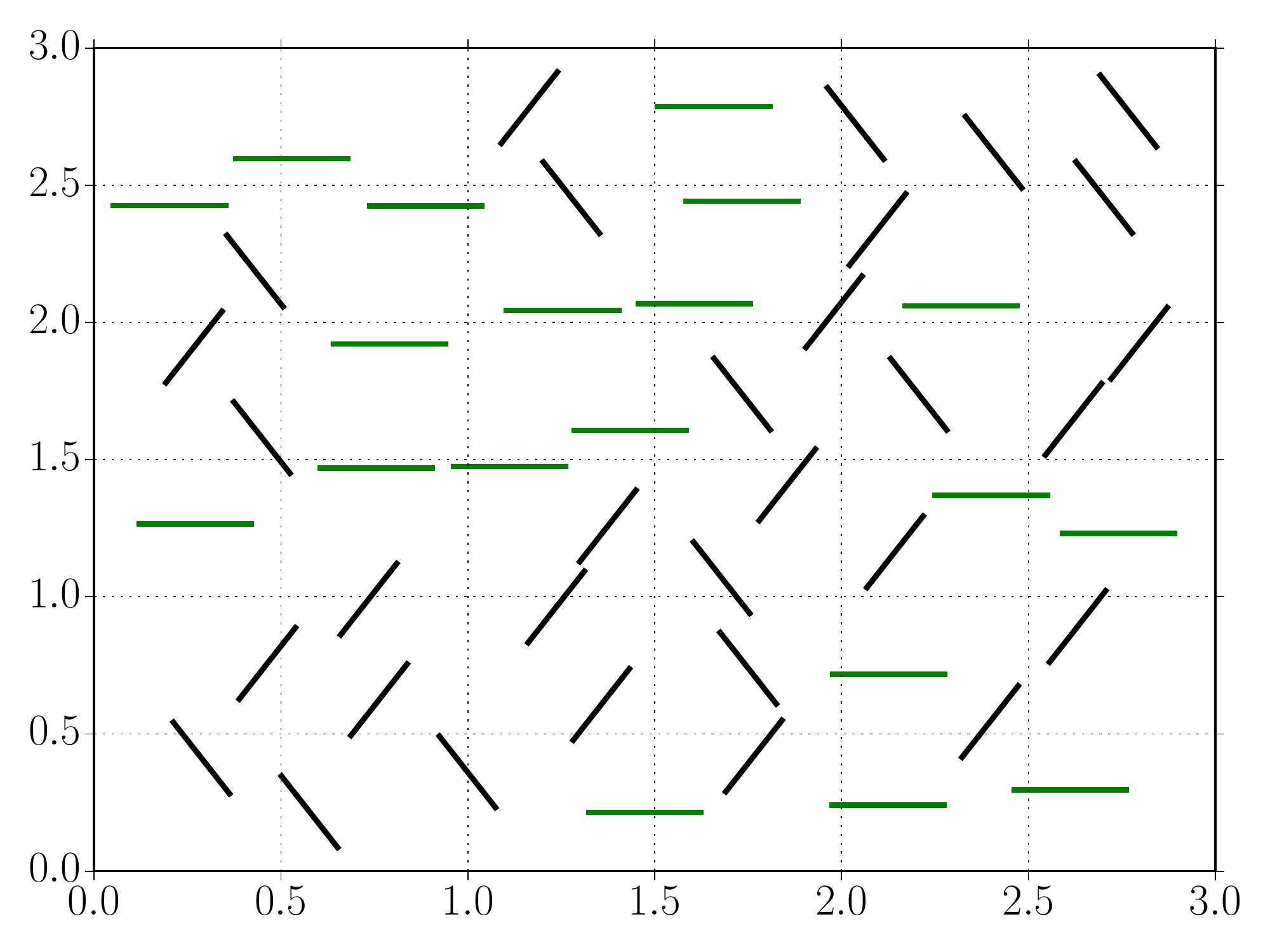}} \\
  \subfigure[A failure path prediction]
    {\includegraphics[width = 0.375\textwidth]
    {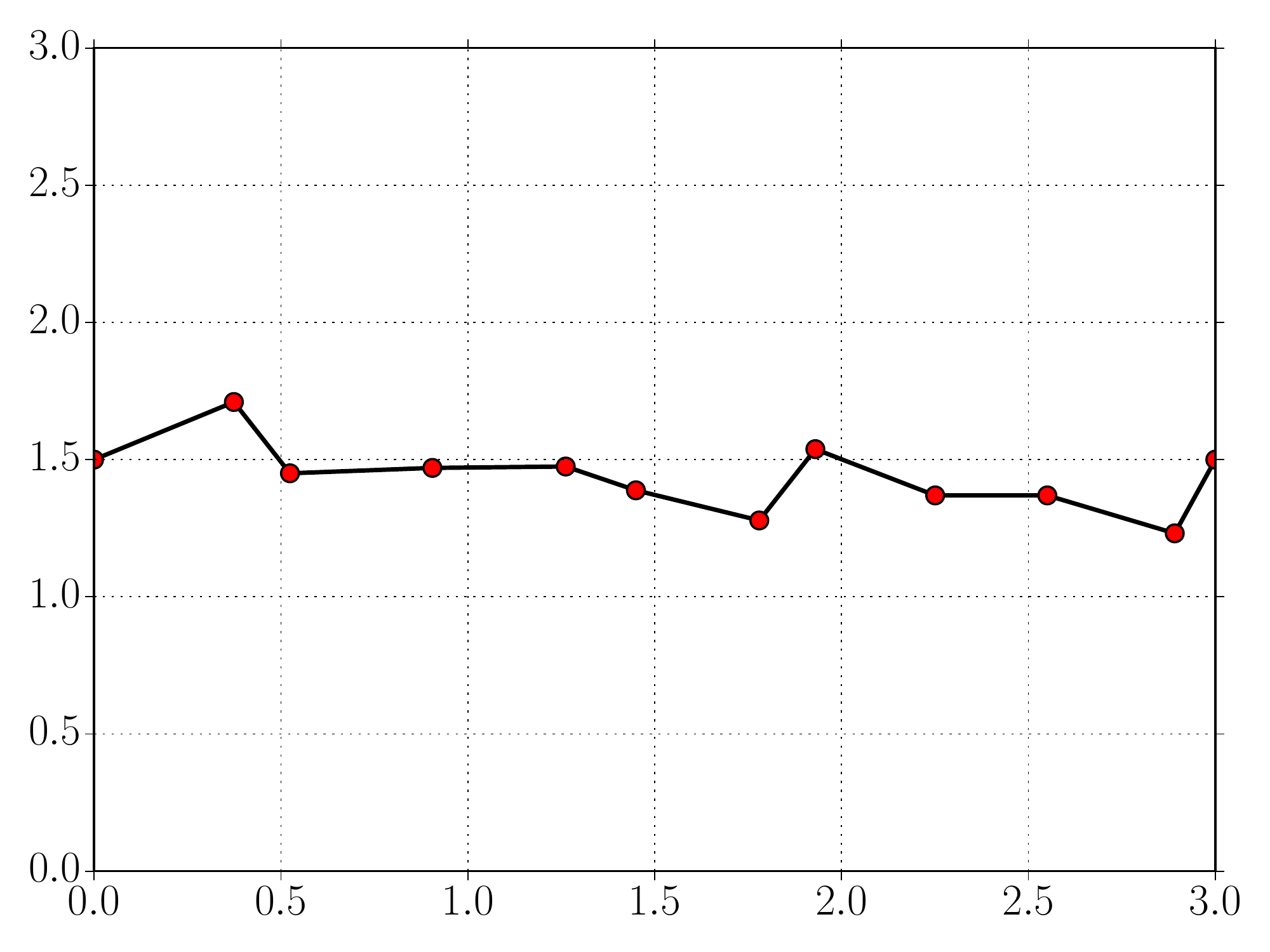}}
  \subfigure[HOSS simulation]
    {\includegraphics[width = 0.29\textwidth]
    {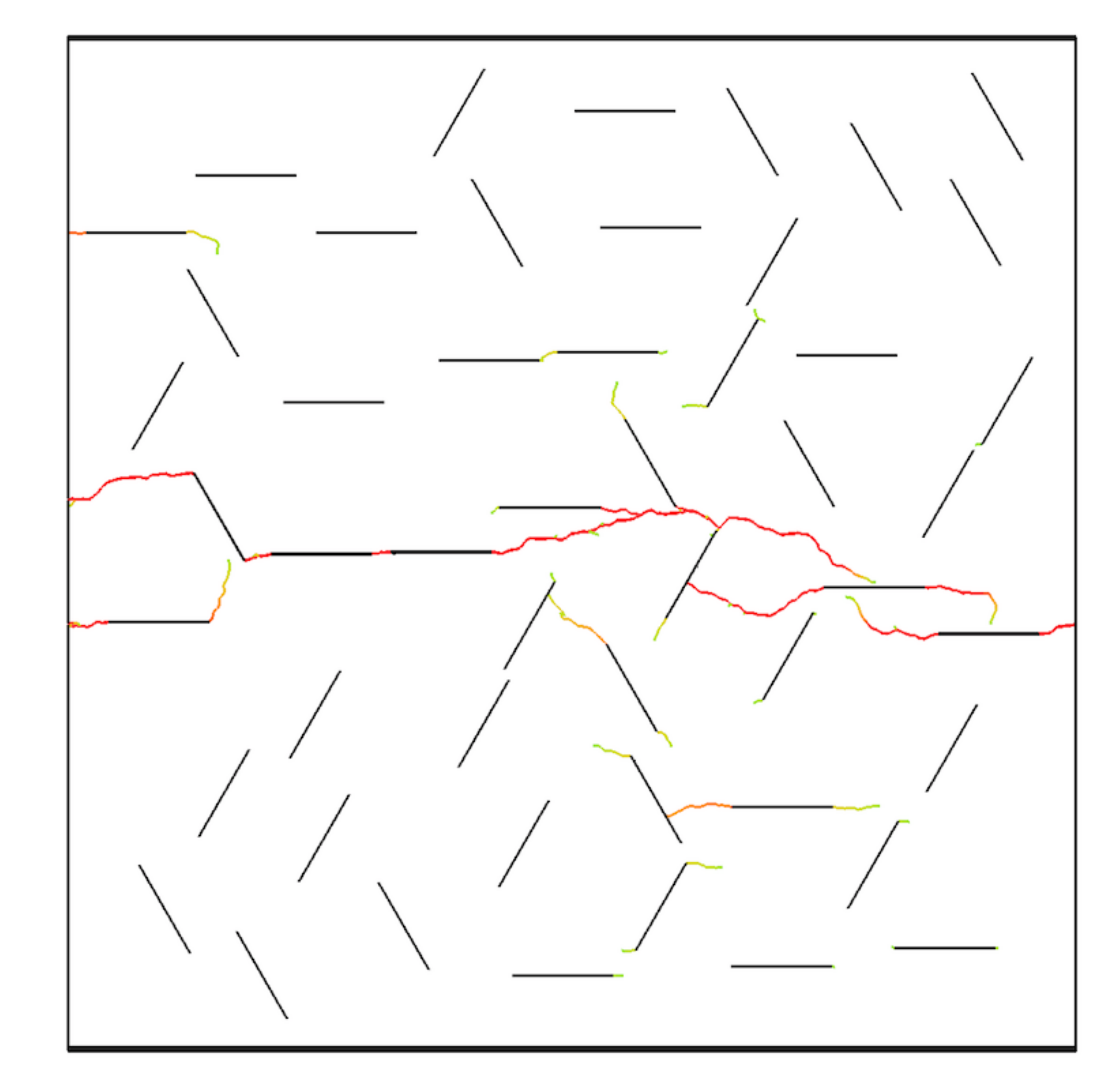}}
  \caption{\textsf{\textbf{50 crack problem:}}~The above figures describe the application of the proposed method for estimating failure paths for 50 crack problem.
  \label{Fig:NFPZ_50_Cracks}}
\end{figure}

Fig.~\ref{Fig:NFPZ_50_Cracks} shows the failure path prediction for an initial configuration with 50 preexisting cracks.
This figure describes the application of the proposed method for estimating failure paths for larger number of cracks in the domain.
The process to predict failure paths is the same as the 20 crack scenario. 
Fig.~\ref{Fig:NFPZ_50_Cracks}(b) shows a failure path prediction, which agrees with HOSS results (see Fig.~\ref{Fig:NFPZ_50_Cracks}(c)).

\begin{table}
  \centering
	\caption{Summary of the proposed method to estimate failure paths.
	  Among various failure paths provided by the proposed method, if a failure path exactly matches the HOSS results, then we consider that as an accurate prediction.
	  Meaning that, all the initial cracks or their crack tips identified by the proposed method fall within the dominant HOSS fracture path.
	  If the proposed method predicts 50\% of the initial cracks or their crack tips to be within the HOSS failure path, then we consider it as a reasonable prediction.
    A non-matching failure path is a scenario where the prediction of the proposed method is less than 50\% match.
	\label{Tab:NFPZ_Accuracy}}
	\begin{tabular}{|c|c|c|} \hline
	  \multirow{2}{*}{{\small Scenario description}} & 
	  \multirow{2}{*}{{\small Number of simulations}} & 
	  \multirow{2}{*}{{\small \% of failure path match}}\\
	  & & \\ \hline\hline
	  {\small Accurate prediction of failure path} & {\small 43}  & {\small 100\% match} \\ 
	  {\small Reasonable prediction of failure path} & {\small 75} & {\small $>$ 50\% match} \\
	  {\small Non-matching failure paths} & {\small 72} & {\small $<$ 50\% match} \\
	  \hline \hline
	  {\small Total} & {\small 190} & \\
	  \hline
	\end{tabular}
\end{table}

\subsection{Estimating damage along the failure paths}
\label{SubSec:S3_Damage_Statistics}
\begin{figure}
  \centering
  \subfigure[Damage evolution model:~Prediction on unseen data]
    {\includegraphics[width = 0.485\textwidth]
    {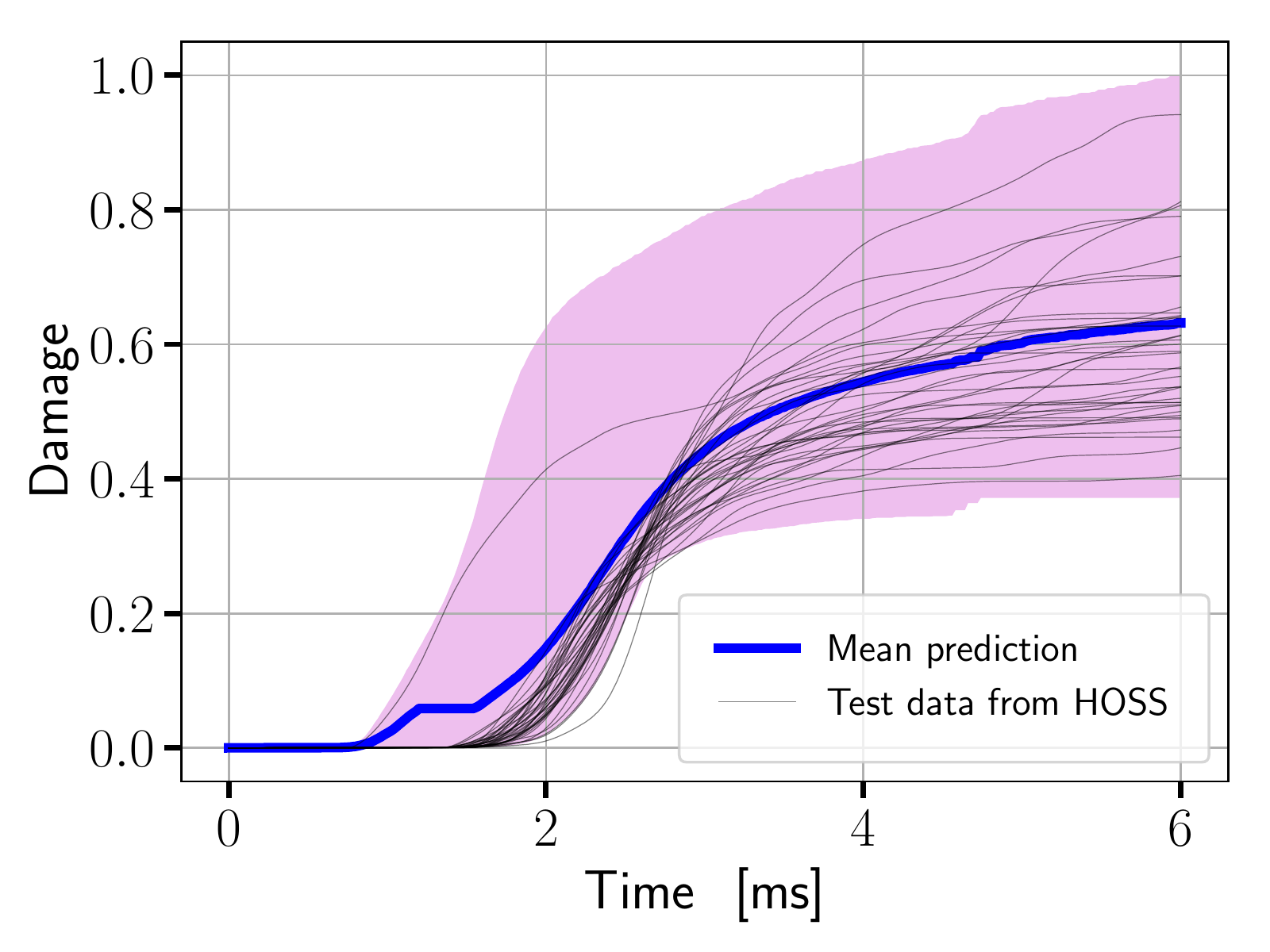}}
  \subfigure[Empirical coverage of proposed damage model]
    {\includegraphics[width = 0.485\textwidth]
    {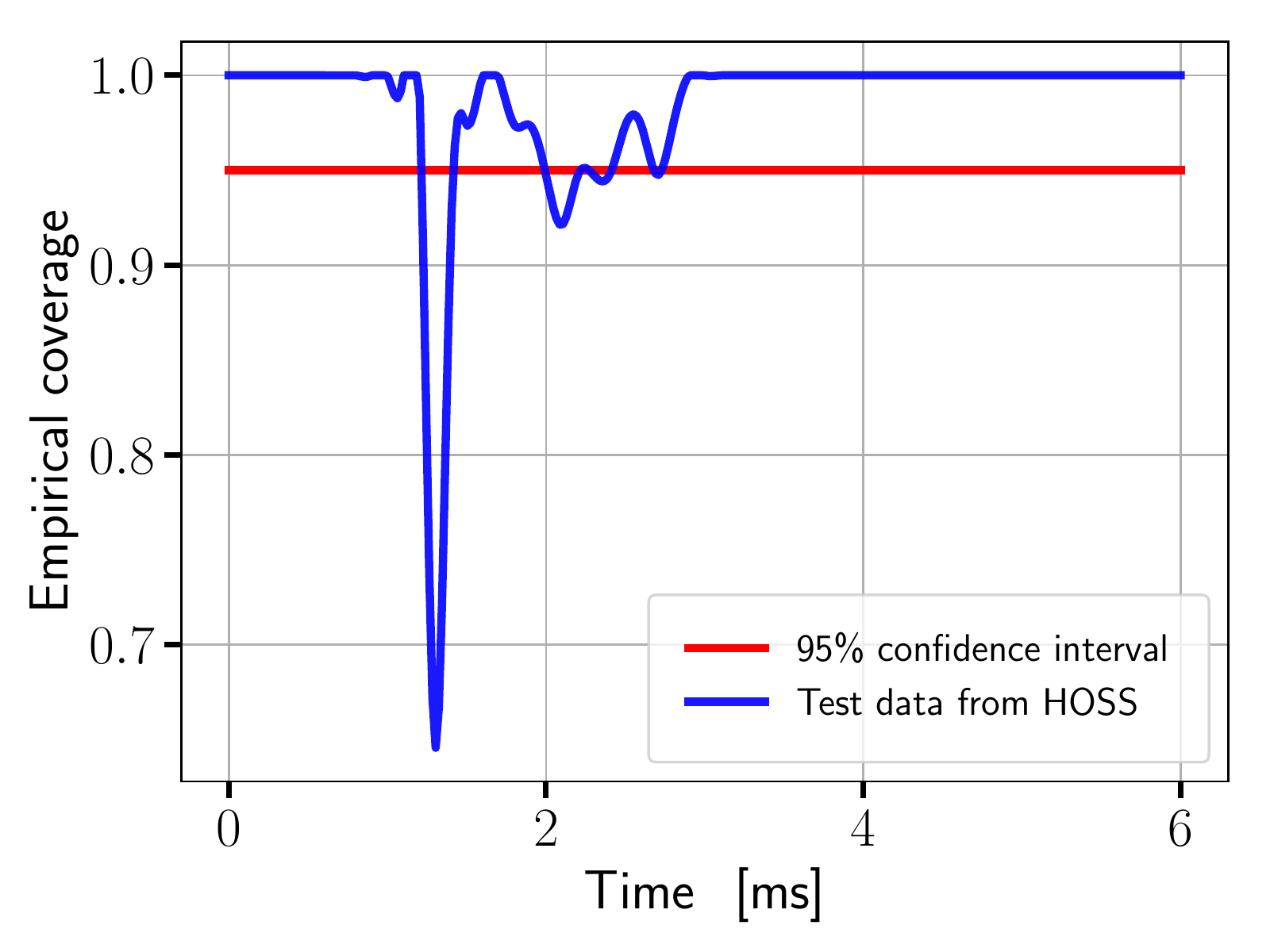}}
  \caption{\textsf{\textbf{Damage statistics:}}~The above figure provide the estimated damage and empirical coverage on data from 40 unseen HOSS simulations.
    From these two figures, it is clear that for most of the times we have more than 90\% prediction accuracy.
  \label{Fig:Damage_Stats}}
\end{figure}

Fig.~\ref{Fig:Damage_Stats}(a) shows the estimated damage evolution vs time, confidence interval, and prediction on unseen data.
In this figure, the mean estimated damage $\overline{D^{(t)}_{M}}$ is shown with a blue line. 
The magenta region shows the damage confidence interval $\left(D^{(t)}_{L}, D^{(t)}_{U}\right)$, which is constructed based on the method discussed in Sec.~\ref{Sec:S4_DynGraph_Damage}.
The test data from HOSS simulations are shown in gray lines.
Fig.~\ref{Fig:Damage_Stats}(a) shows that our confidence interval captures the damage from the test set. 

Fig.~\ref{Fig:Damage_Stats}(b) provides the corresponding empirical coverage of the proposed surrogate model for damage.  
The blue line corresponds to the empirical coverage on the test dataset from HOSS.
The red line corresponds to 95\% confidence interval.
As discussed in Sec.~\ref{Sec:S4_DynGraph_Damage}, empirical coverage is a measure of how well our predicted confidence interval captures unseen data.
From this figure, we see that we are able to capture more than $90\%$ of the test data over time except for a small region around time 1.75 ms where we under cover (this is because the cracks along the failure path start to grow around this time).
This prediction accuracy ($> 90\%$) implies that the proposed method can be used to develop probability density functions for damage evolution for a representative volume element.
As the damage evolution model is trained based on HOSS simulations, it contains crack interaction effects.
During the upscaling process, this crack interaction information is contained in the effective modulii.
This implies that the material response we obtain at continuum scale includes crack interactions at micro-scale, which are typically lost if one employs traditional mechanistic or homogenization approaches \cite{graham2013upscaling,paliwal2008interacting,krajcinovic1995some,krajcinovic1987micromechanics}.
%

\section{CONCLUSIONS}
\label{Sec:S6_DynGraph_Conclusions}
Predicting damage evolution and when failure occurs is important to accurately predict the overall material response \cite{chillara2017thermodynamic,2012_Freiman_Mecholsky}.
This includes accounting for degraded material properties as damage accumulates and when and where failure of the specimen or part occurs, for example, which elements or cells will fail first. 
Providing such information is of great interest to the fracture and infrastructure maintenance communities.
In addition, estimating likely failure paths and accumulated damage along the failure path is an important aspect for upscaling micro-scale models to macro-scale models.
FDEM models and multi-physics numerical tools like HOSS are highly-accurate to simulate various stages of brittle fracture propagation.
However, HOSS is data-intensive and computationally expensive.
Moreover, scaling to larger and more complex problems is often computationally prohibitive with computational tools such as HOSS.
This is due to the difference in length scales associated with small fractures and bulk material sizes.
In this paper, we provided algorithms to address these aspects that are $\approx \mathcal{O} (10^6)$ faster than high-fidelity simulations to estimate key QoIs for brittle fracture such as failure paths, failure zones, and damage along failure.
Hence, the proposed methodology is ideal for usage in comprehensive uncertainty quantification studies.

The proposed failure path method was compared against high-fidelity HOSS simulations.
From this comparison, we found that out of 190 simulations, the proposed method predicted failure paths of 118 simulations with
reasonable accuracy (greater than 50\% of initial cracks in the failure path) and zone of failure of 143 simulations with 100\% accuracy.
Damage along the failure path was estimated using a non-parametric approach.
The damage evolution model was obtained by constructing a 95\% confidence interval over time. 
150 HOSS simulations were used to develop the damage evolution model and the remaining 40 simulations are used for testing.
Bootstrapping was used to estimate lower and upper bounds for damage.
Empirical coverage was used as a metric to understand the accuracy of the proposed damage evolution model.
Empirical coverage tells us the fraction of test cases that falls within our estimated confidence interval at anytime. 
From empirical coverage results, it can be concluded that our confidence interval captures the damage from the test dataset with an accuracy greater than 90\%.
Damage is the key QoI for upscaling HOSS information to continuum codes such as FLAG.
This is because continuum-scale models need effective moduli, which is a function of damage accumulated over time at micro-scale.
Since the discrete crack network cannot be accounted for in continuum models, exact locations of the failure pathways becomes less important.
The proposed damage model which takes into account the crack interaction effects provides such micro-scale information with high accuracy for simulating damage at larger-scales.
The limitation of the proposed methods is that it is applicable only for Mode I failure at low-strain rates.
Extensions to higher-strain rates and other modes of failure will be considered in our future works.
Lastly, our method can be coupled with other machine learning and graph-based methods \cite{sundararaghavan2017microfract,2018_Hunter_etal_arXiv,khodabakhshi2016grafea,lu2011link} for increased accuracy of failure paths prediction, which is our future work.

\section*{ACKNOWLEDGMENTS}
The authors thank the support of the LANL Laboratory Directed Research and Development Directed Research Award 20170103DR. 
MKM and SK authors also thank the support of the LANL Laboratory Directed Research and Development Early Career Award 20150693ECR. 
MKM gratefully acknowledges the support of LANL Chick-Keller Postdoctoral Fellowship through Center for Space and Earth Sciences (CSES).
Authors thank the LANL Institutional Computing program for their support in generating data used in this work.
The authors also thank Bryan Moore for providing HOSS simulation datasets.

\bibliographystyle{unsrt}
\bibliography{Master_References/Books,Master_References/Master_References}
\newpage


\end{document}